\documentclass[longbibliography,noeprint,noshowpacs,nopreprintnumbers,pra,twocolumn,showpacs,superscriptaddress,amsmath,amssymb,citeautoscript,aps,10pt]{revtex4-2}
\usepackage{graphicx}
\usepackage{dcolumn}
\usepackage{color}
\usepackage{bm}
\usepackage{chngcntr}
\usepackage[hidelinks]{hyperref}
\hypersetup{
    colorlinks,
    citecolor=blue,
    filecolor=blue,
    linkcolor=blue,
    urlcolor=blue
}
\graphicspath{{Figures/}{./Figures/}{.}}

\DeclareMathOperator{\tr}{tr}
\DeclareMathOperator{\C}{\mathcal{C}}
\DeclareMathOperator{\Hm}{\mathcal{H}}

\begin{document}

\title{Dynamical bulk boundary correspondence and dynamical quantum phase transitions in higher order topological insulators}

\author{T. Mas{\l}owski}
\affiliation{The Faculty of Mathematics and Applied Physics, Rzesz\'ow University of Technology, al.~Powsta\'nc\'ow Warszawy 6, 35-959 Rzesz\'ow, Poland}
\author{N.~Sedlmayr}
\email[e-mail:]{sedlmayr@umcs.pl}
\affiliation{Institute of Physics, M. Curie-Sk\l{}odowska University, 20-031 Lublin, Poland}

\date{\today}

\begin{abstract}
Dynamical quantum phase transitions occur in dynamically evolving quantum systems when non-analyticities occur at critical times in the return rate, a dynamical analogue of the free energy. This extension of the concept of phase transitions can be brought into contact with another, namely that of topological phase transitions in which the phase transition is marked by a change in a topological invariant. Following a quantum quench dynamical quantum phase transitions can happen in topological matter, a fact which has already been explored in one dimensional topological insulators and in two dimensional Chern insulators. Additionally in one dimensional systems a dynamical bulk boundary correspondence has been seen, related to the periodic appearance of zero modes of the Loschmidt echo itself. Here we extend both of these concepts to two dimensional higher order topological matter, in which the topologically protected boundary modes are corner modes. We consider a minimal model which encompasses all possible forms of higher order topology in two dimensional topological band structures. We find that DQPTs can still occur, and can occur for quenches which cross both bulk and boundary gap closings. Furthermore a dynamical bulk boundary correspondence is also found, which takes a different form to that in one dimension.
\end{abstract}

\maketitle

\section{Introduction}\label{sec:intro}

A dynamical quantum phase transition (DQPT) is said to occur when non-analyticities appear at critical {\it times} in the return rate, a measure of the overlap between a time evolved and initial state~\cite{Heyl2013,Heyl2018a,Sedlmayr2019a}. More specifically the return rate is proportional to the log of the magnitude of the overlap itself, called the Loschmidt echo. This has a clear analogy to non-analyticities in the free energy which appear at critical parameter values across quantum phase transitions. The paradigmatic case is for quenches in which the initial state is the ground state of one Hamiltonian, and the system is then time evolved with a different Hamiltonian. Generalisations of DQPTs have been made to mixed states, finite temperatures, and open or dissipative systems~\cite{Mera2017,Sedlmayr2018b,Bhattacharya2017a,Heyl2017,Abeling2016,Lang2018,Lang2018a,Kyaw2020,Starchl2022,Naji2022,Kawabata2022}, with somewhat mixed results. Whether the DQPTs survive depends on details both of the models studied and the particular generalization of the return rate that is used. In the relatively simple model that was first studied there appeared to be a direct connection between the existence of DQPTs and the equilibrium phase diagram: DQPTs only occurred if the quench crossed an equilibrium phase boundary. However it was realised son after that such a general one-to-one correspondence between the equilibrium phase diagram and DQPTs does not exist~\cite{Vajna2014,Andraschko2014,Vajna2015,Karrasch2017,Jafari2017,Jafari2017a,Cheraghi2018,Jafari2019,Wrzesniewski2022}. DQPTs can therefore be said to offer real insight into non-equilibrium phenomena with the advantage that it is in a simple and controlled way. Furthermore interesting connections have been found between DQPTs and several other phenomena such as the entanglement entropy~\cite{Sedlmayr2018}, string order parameters~\cite{Uhrich2020}, the characteristic function of work~\cite{Abeling2016,Talkner2007}, and crossovers in the quasiparticle spectra~\cite{Halimeh2018}.

Although focus began on simple two band one dimensional models, this has been extended to multi-band models~\cite{Huang2016,Jafari2019,Mendl2019,Maslowski2020}, and two dimensional systems~\cite{Vajna2015,DeNicola2022,Hashizume2022,Brange2022}. In the case of two dimensional systems there is an additional complication: the existence of extended lines of critical times with finite length, rather than critical points. In this case the DQPT manifests itself more directly in the time derivative of the return rate~\cite{Vajna2015}. A large amount of theoretical work has followed~\cite{Heyl2013,Karrasch2013,Sharma2014,Heyl2014,Heyl2015,Sharma2015,Halimeh2017,Homrighausen2017,Halimeh2018,Shpielberg2018,Zunkovic2018,Yang2019,Srivastav2019,Huang2019,Gurarie2019,Abdi2019,Puebla2020,Link2020,Sun2020,Rylands2020,Trapin2021,Yu2021,Halimeh2021,Halimeh2021a,DeNicola2021,Cheraghi2021,Cao2021,Bandyopadhyay2021}.  Experimentally a variety of approaches have been used to realise DQPTs including ion traps, cold atoms, and quantum simulator platforms~\cite{Jurcevic2017,Flaschner2018,Zhang2017b,Guo2019,Smale2019,Nie2020,Tian2020}. 

DQPTs have also been shown to exist in a variety of topological models~\cite{Vajna2015,Schmitt2015,Jafari2016,Jafari2017a,Sedlmayr2018,Hagymasi2019,Jafari2018,Zache2019,Maslowski2020,Porta2020,Mishra2020,Uhrich2020,Okugawa2021,Sadrzadeh2021}. Crossing a topological phase boundary with a quench often results in DQPTs, but is neither necessary nor sufficient.  Previously focus was on topological band structures which can be characterised by topological indices such as the Zak-Berry phase~\cite{Zak1989} or Chern number~\cite{Hasan2010}. Of great importance for the topological classification are the symmetries of the Hamiltonians~\cite{Schnyder2009,Ryu2010}, an idea which can be extended to crystalline symmetries~\cite{Fu2011,Xu2012,Zhang2013a,Liu2014,Shiozaki2014}. One of the most interesting consequences of topological band structures is of course the bulk-boundary correspondence~\cite{Hasan2010,Teo2010} and the existence of topologically protected boundary modes with one dimension lower than the bulk. In a higher order topological insulator (HOTI) the dimension of the edge modes can be two or more lower than the bulk dimension~\cite{Volovik2010,Sitte2012,Zhang2013b,Benalcazar2017,Benalcazar2017a,Langbehn2017,Song2017,Schindler2018,Fang2019,Trifunovic2018,Trifunovic2021,Xie2021}, i.e.\ one can have modes at the corners of two dimensional and three dimensional topological matter, or along the hinges of a three dimensional crystal. As an example this can be loosely understood as resulting from breaking a crystalline symmetry along the one dimensional edge of a two dimensional system which would otherwise have boundary modes. These one dimensional edge modes become gapped and can in turn lead to one dimensional corner modes where they meet.

A natural question to ask in this context is if can one have dynamical order parameters~\cite{Budich2016,Sharma2016,Heyl2017,Bhattacharya2017a,Dutta2017} and a dynamical bulk-boundary correspondence~\cite{Sedlmayr2018,Sedlmayr2019a,Maslowski2020} related to the dynamical quantum phase transitions. A dynamical order parameter can be introduced via the phase of the complex Loschmidt echo. DQPTs are caused by zeroes of the Loschmidt echo, which results in a phase jump in the phase, and by extension in the dynamical order parameter. This does not appear to entail any further information than what is already contained in the Loschmidt echo, but may be another method of measuring the DQPTs~\cite{Wang2019e}. In contrast the dynamical bulk-boundary correspondence considers boundary contributions to the Loschmidt echo or return rate which, as boundary contributions, do not appear in these quantities in the thermodynamic limit. It is found that, depending in the topology of the time evolving Hamiltonian, the Loschmidt echo develops zero modes which periodically appear and disappear at the critical times. These result in characteristic plateaus forming in the boundary contribution to the return rate.

In this article we investigate DQPTs in two dimensional HOTIs with corner modes. We consider both intrinsic and extrinsic cases with both two and four corner modes present. For this purpose we introduce a model which encompasses all these phases based on the Benalcazar-Bernevig-Hughes (BBH) model~\cite{Benalcazar2017,Benalcazar2017a}. We then extend the concept of the dynamical bulk-boundary correspondence to DQPTs in HOTIs. In these HOTIs the topological phase can change not only via the bulk gap closing, but also by closing the edge gap without the bulk gap closing. One can also change the phase by breaking symmetries. For quenches which cross both types of gap closing we find DQPTs, however quenches which break or restore symmetries, without crossing gap closings do not have DQPTs. For the relatively simple model here we don't find DQPTs for quenches within a phase. As for the one-dimensional cases previously studied we find that zeroes of the Loschmidt echo occur between critical times when the time evolving Hamiltonian is topologically non-trivial, though the pattern is more complicated than for one dimensional topological systems. None of these results depends qualitatively on whether we consider an extrinsic or an intrinsic HOTI.

This article is organised as follows. In section \ref{sec:mods} we introduce our generalized BBH model along with exemplary spectra and its phase diagram. In Sec.~\ref{sec:los} we introduce the definitions of the Loschmidt echo and return rate and the details of the quenches we will focus on. Sec.~\ref{sec:dqpts} presents results for the Fisher zeroes and DQPTs for a variety of the quenches we explore. In Sec.~\ref{sec:dbb} this is then related to the dynamical bulk-boundary correspondence and in Sec.~\ref{sec:dis} we conclude.

\section{Model}\label{sec:mods}

In general there are several types of behaviour a two dimensional HOTI can display. It can have either two or four corner modes present and additionally the topology may be though of as extrinsic or intrinsic. For an intrinsic HOTI the topology is protected by a bulk crystalline symmetry which is absent for the extrinsic case. Here we introduce a minimal four band model which includes all of these possibilities. Let us consider the Hamiltonian
\begin{equation}
\Hm_{m,n,n'}
= J\vec{\Gamma} \cdot\vec{d}_{m,n,n'}\,,
\end{equation}
where $\vec{\Gamma}$ is a vector containing eight $4\times4$ matrices. The matrices are given by $\Gamma_k=-\tau_2\sigma_k$ and $\Gamma_{4+k}=-\tau_1\sigma_k$ for $k=1,2,3$, and by $\Gamma_4=\tau_1\sigma_0$ and $\Gamma_8=\tau_2\sigma_0$. The momentum dependent vector is
\begin{equation}  
  \vec{d}_{m,n,n'} =
    \begin{pmatrix}
       (1 - n)\sin k_y\\
       m+(1 - n)\cos k_y\\
       \sin k_x + n'\,\sin k_y\\
       m+\cos k_x + n\,\cos k_y\\
       n\,\cos k_y \\ 
      -n\,\sin k_y\\
      -n'\,\cos k_y \\
      -n\,\sin k_y
    \end{pmatrix}\,.
\end{equation}
$J$ is an overall energy scale of the hopping terms and we will set everywhere $J=1$ and $\hbar=1$. For $n=n'=0$ we reproduce the BBH Hamiltonian with a possible 4 corner states \cite{Benalcazar2017,Benalcazar2017a}. We will consider two variants of this general Hamiltonian. First we have $n'=0$ which as we will see is an intrinsic HOTI with two or four corner modes. Second we have $n'\neq0$ which is an extrinsic HOTI with two or four corner modes~\cite{Trifunovic2021}, and we focus particularly on the case $n=n'$. Throughout this paper intrinsic will be used specifically to refer to $n'=0$ and extrinsic to $n=n'$.

\begin{figure}[t!]
\includegraphics[width=0.95\columnwidth]{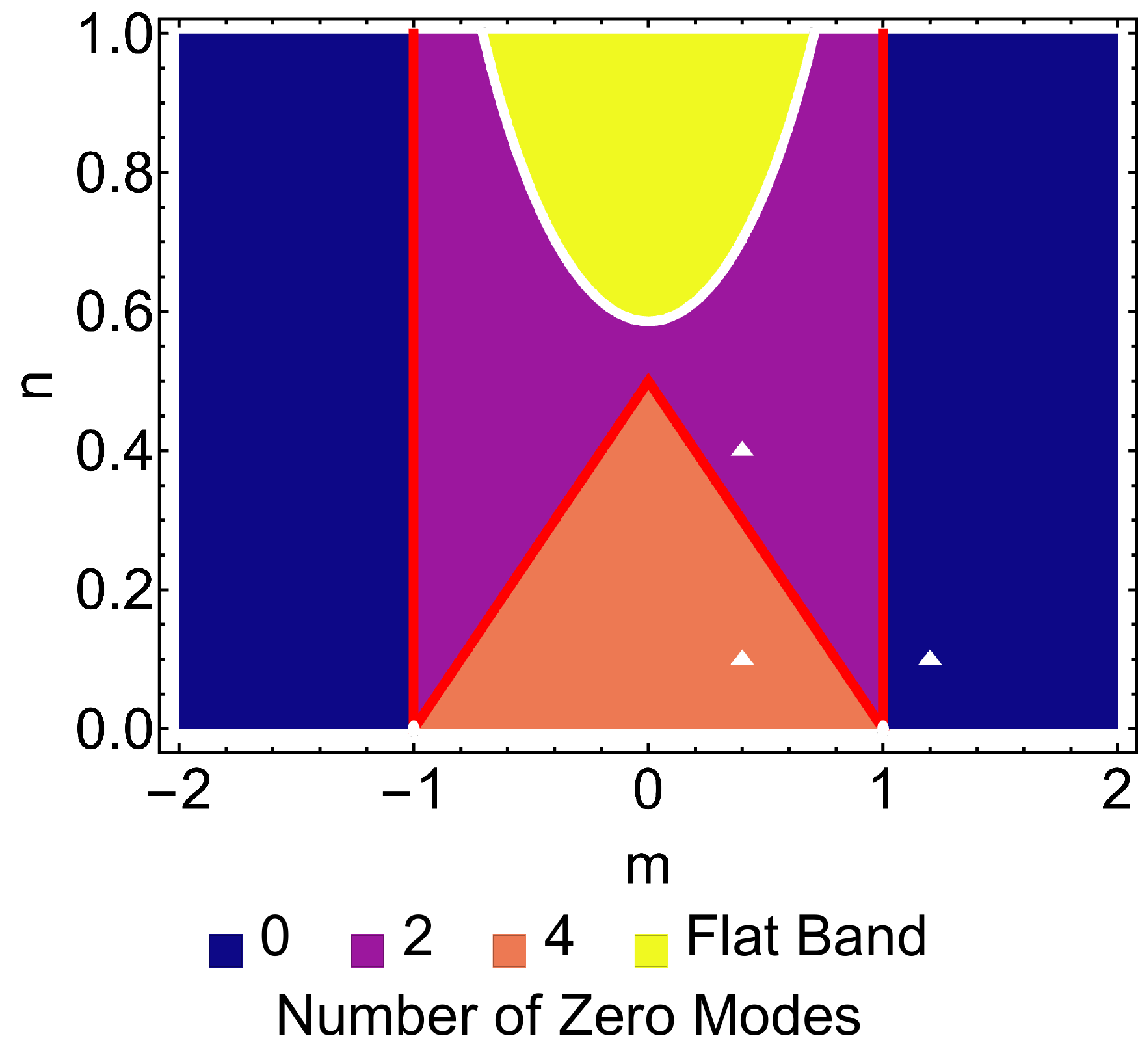}
\caption{The topological phase diagram for the intrinsic case $n'=0$. The number of corner modes is given in the legend. The yellow region marked as having a flat band has flat bands between bulk gap closing points, see appendix \ref{app:mod} for more details. In this analysis we focus only on quenches between regions with 0, 2, or 4 zero modes, not the flat band region. The white lines are bulk gap closing topological phase transitions and the red lines show edge gap closing topological phase transitions. Edge gap closing always means the otherwise gapped edge modes which exist {\it along} the $y$-direction. The white triangles show the points used in the quenches.} 
\label{fig:phasei}
\end{figure}

This model has a global particle hole symmetry, $\mathcal{C}=\tau_3\times\sigma_0\hat K$, satisfying $\{\mathcal{C},\Hm_{m,n,n'}\}=0$ and $\mathcal{C}^2=1$. It also has a ``time-reversal'' symmetry $\mathcal{T}=\tau_0\times\sigma_0\hat K$ satisfying $\{\mathcal{T},\Hm_{m,n,n'}\}=0$ and $\mathcal{T}^2=1$. $\hat K$ is charge conjugation. There are also crystalline symmetries present. For $n=0$ we find the mirror symmetries~\cite{Benalcazar2017a,Trifunovic2021}
\begin{equation}
    \mathcal{U}_y\Hm_{m,0,0}(-k_x,k_y)\mathcal{U}_y^\dagger=\Hm_{m,0,0}(k_x,k_y)
\end{equation}
and
\begin{equation}
    \mathcal{U}_x\Hm_{m,0,0}(k_x,-k_y)\mathcal{U}_x^\dagger=\Hm_{m,0,0}(k_x,k_y)
\end{equation}
where $\mathcal{U}_y=\tau_1\sigma_3$ and $\mathcal{U}_x=\tau_1\sigma_1$. We also have a four fold rotational symmetry
\begin{equation}
    \mathcal{U}_4\Hm_{m,0,0}(-k_y,k_x)\mathcal{U}_4^\dagger=\Hm_{m,0,0}(k_x,k_y)
\end{equation}
with
\begin{equation}
    \mathcal{U}_4=\begin{pmatrix}
    0&0&1&0\\
    0&0&0&1\\
    0&-1&0&0\\
    1&0&0&0
    \end{pmatrix}
\end{equation}
and $\mathcal{U}_4^4=-1$. By combining the rotation and mirror symmetries it is therefore also possible to write the following mirror symmetries:
\begin{equation}
    \mathcal{U}_{x=-y}\Hm_{m,0,0}(-k_y,-k_x)\mathcal{U}_{x=-y}^\dagger=\Hm_{m,0,0}(k_x,k_y)
\end{equation}
and
\begin{equation}
    \mathcal{U}_{x=y}\Hm_{m,0,0}(k_y,k_x)\mathcal{U}_{x=y}^\dagger=\Hm_{m,0,0}(k_x,k_y)
\end{equation}
where $\mathcal{U}_{x=-y}=\mathcal{U}_{y}\mathcal{U}_4$ and $\mathcal{U}_{x=y}=\mathcal{U}_{x}\mathcal{U}_4$. More details on the symmetry operations and the matrices can be found in appendix \ref{app:mod}.

For $n\neq0$ the crystalline symmetries $\mathcal{U}_{y}$, $\mathcal{U}_{x=-y}$, $\mathcal{U}_{x=y}$, and $\mathcal{U}_{4}$ are broken, leaving only $\mathcal{U}_{y}$ intact. This last one broken by $n'\neq0$. Therefore we find that $\Hm_{m,n,n'\neq0}$ is an extrinsic HOTI and $\Hm_{m,n,0}$ is an intrinsic HOTI each with either 2 or 4 corner modes.

\begin{figure}
\includegraphics[width=0.95\columnwidth]{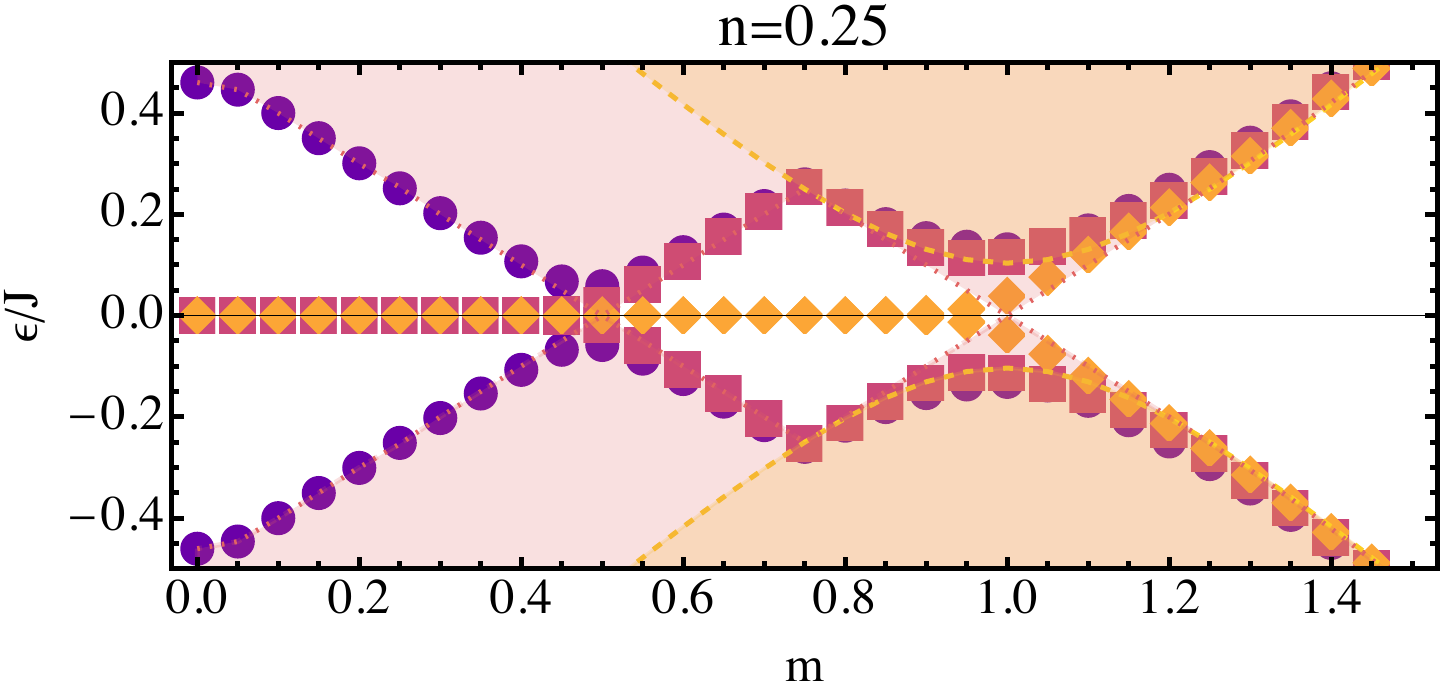}
\includegraphics[width=0.95\columnwidth]{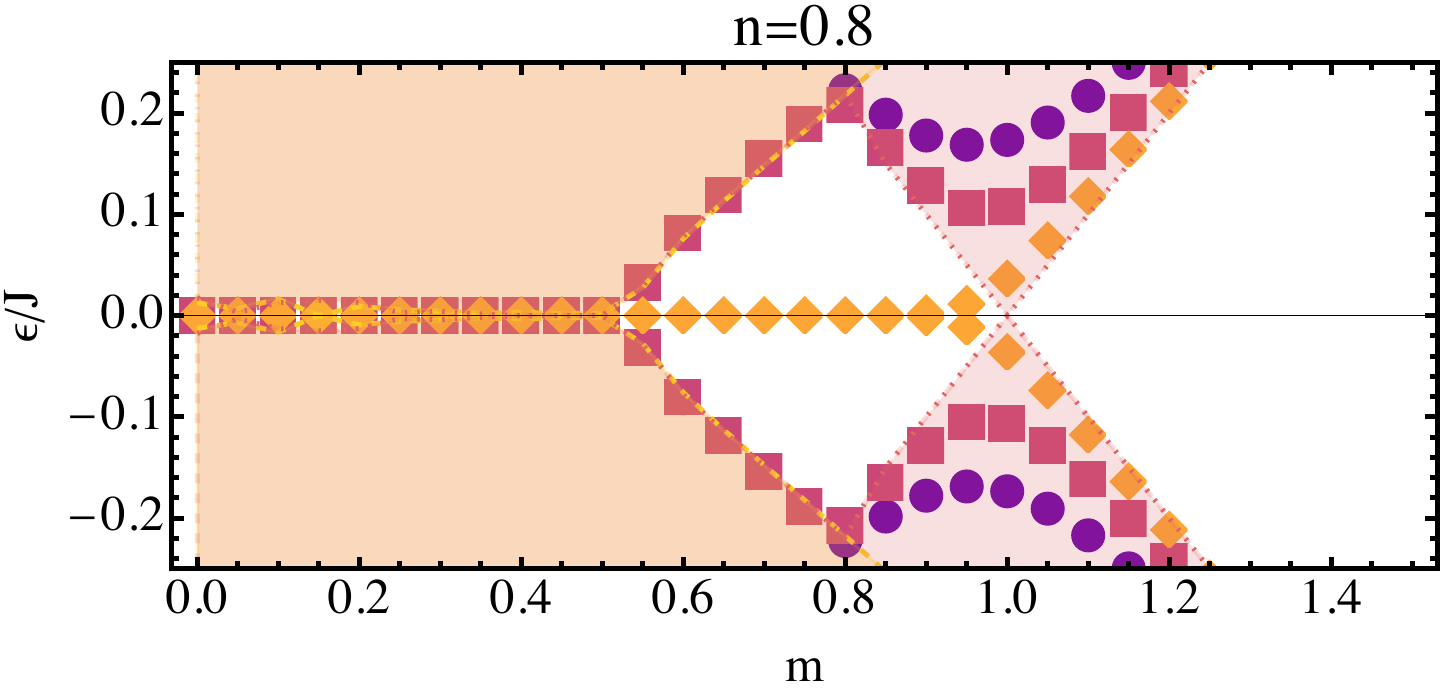}
\caption{Low energy spectra of the intrinsic HOTI through two cuts in the phase diagram. The lowest six energy levels of a nanoflake of size $40\times 40$ are shown. The light pink shaded areas show the one dimensional edge states and the light orange shaded area shows the bulk states.} 
\label{spectrai}
\end{figure}

In Fig.\ \ref{fig:phasei} the intrinsic topological phase diagram is shown, with the positions used for the quenches in the following sections marked. The phases are divided by the edge gap closing lines $|n|+|m|=1/2$ and $|m|=1$; and the bulk gap closing line $(n-2)^2=2(1-m^2)$. The flat band region has flat bands along the y-direction which lie between bulk gap closing, see appendix \ref{app:mod} for examples of the band structure. In Fig.~\ref{spectrai} we show the single particle spectra $\epsilon$ along two cuts through the phase diagram. At $n=0.25$ we see the edge gap closing between four and two corner modes, followed by the edge gap closing to the topologically trivial phase. Although the bulk gap narrows, it does not in fact close at this point. At $n=0.8$ one can see the gapless phase containing flat bands which gives way to the phase with two corner modes, followed by the edge gap closing to the topologically trivial phase.

\begin{figure}[t!]
\includegraphics[width=0.95\columnwidth]{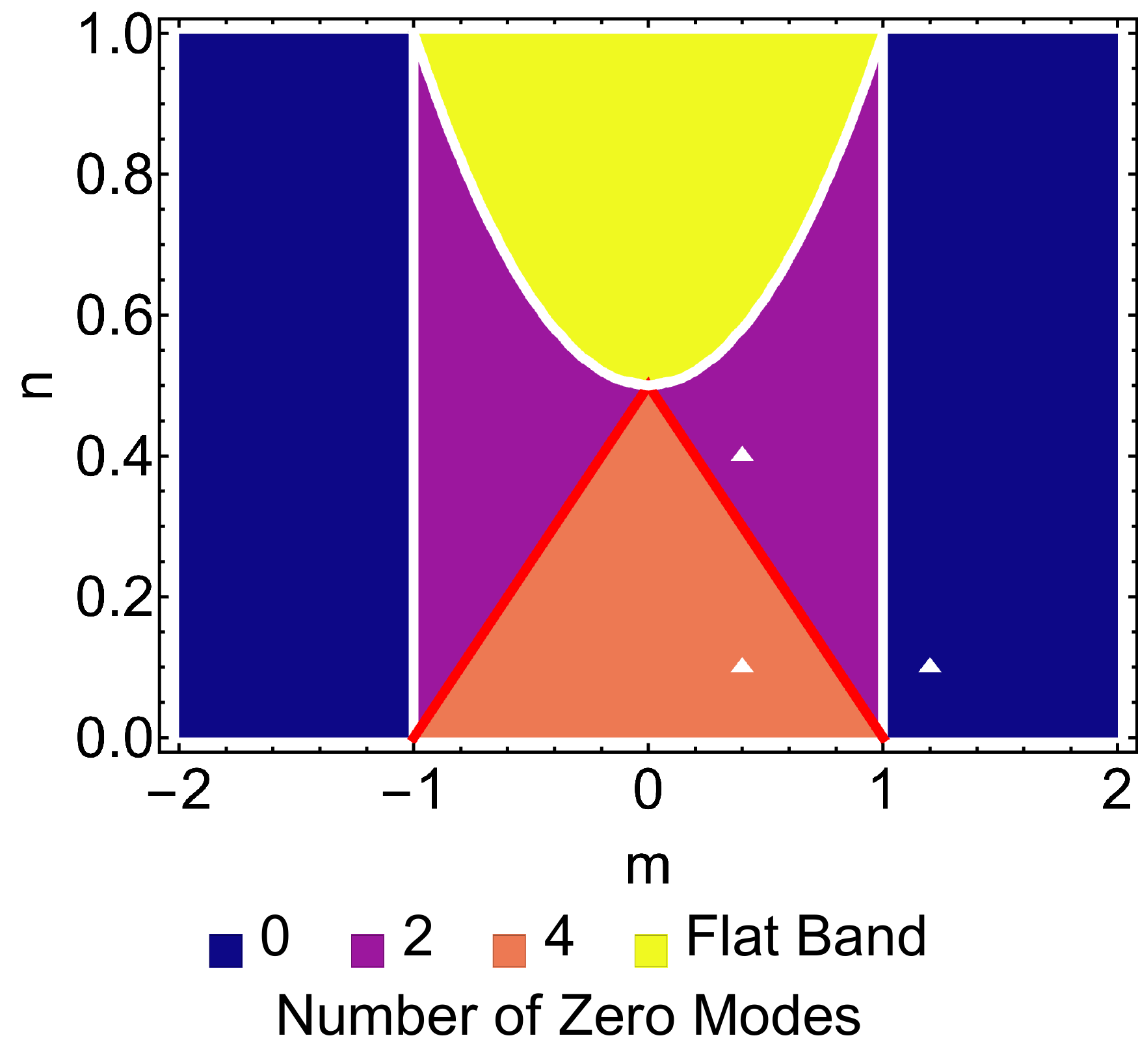}
\caption{The topological phase diagram in an extrinsic case, specifically $n'=n$. The number of corner modes is given in the legend. The yellow region marked as having a flat band is fully gapped in the bulk, see appendix \ref{app:mod} for more details. In this analysis we focus only on quenches between regions with 0,2, or 4 zero modes, not the flat band region. The white lines are bulk gap closing topological phase transitions and the red lines show edge gap closing topological phase transitions. Edge gap closing always means the otherwise gapped edge modes which exist {\it along} the $y$-direction. The white triangles show the points used in the quenches} 
\label{fig:phasee}
\end{figure}

\begin{figure}
\includegraphics[width=0.95\columnwidth]{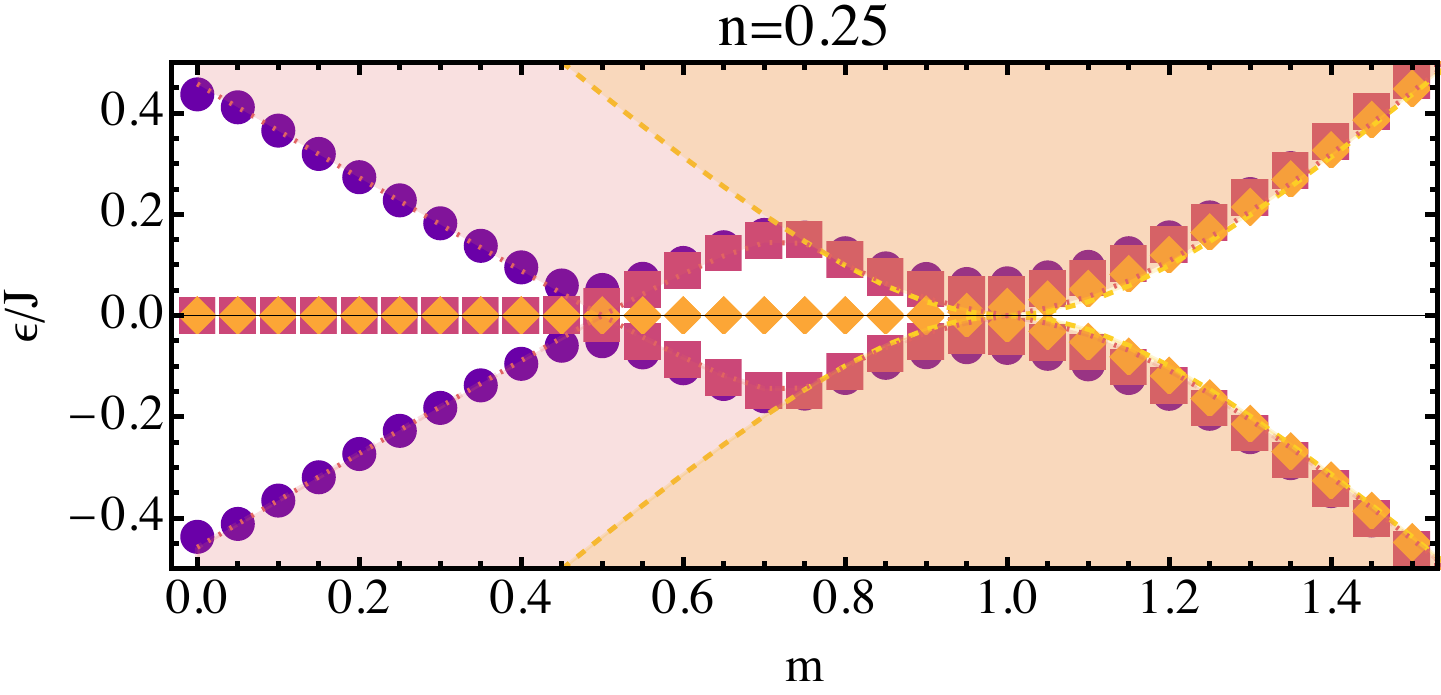}
\includegraphics[width=0.95\columnwidth]{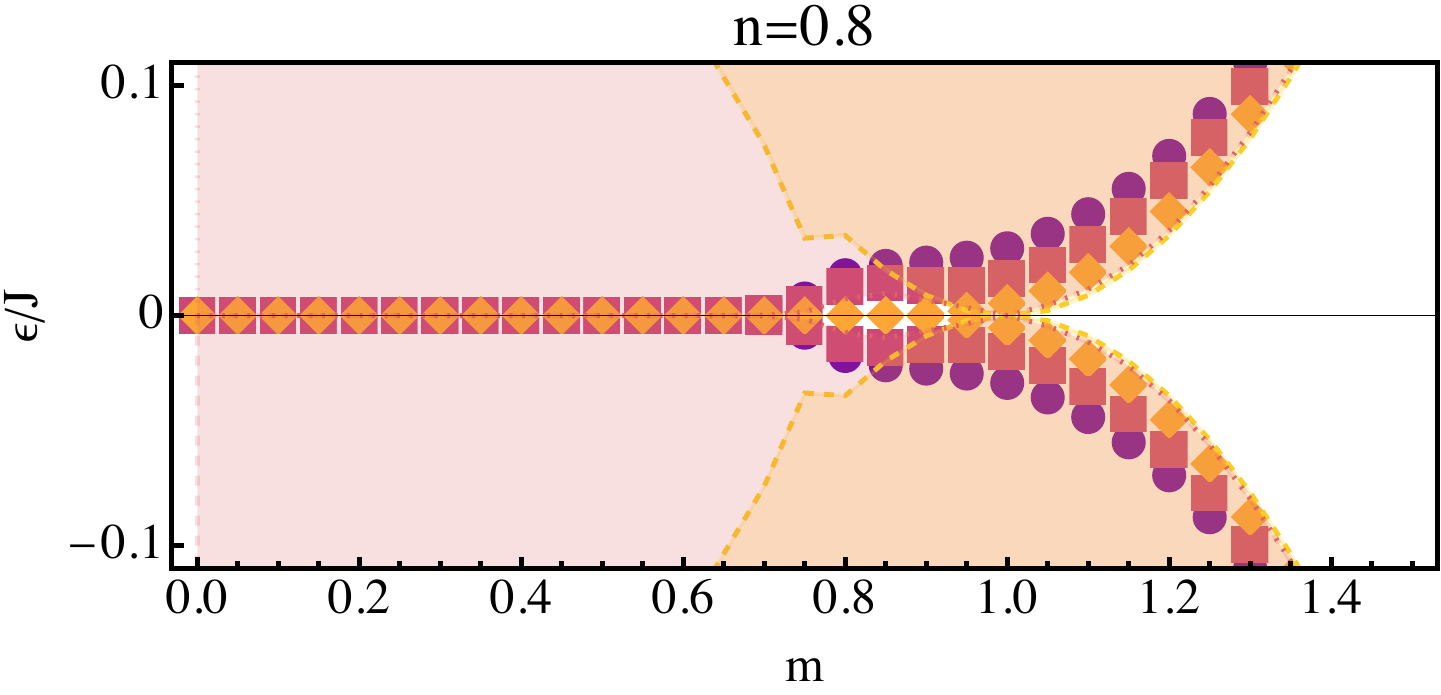}
\caption{Low energy spectra of the extrinsic HOTI through two cuts in the phase diagram. The lowest six energy levels of a nanoflake of size $40\times 40$ are shown. The light pink shaded areas show the one dimensional edge states and the light orange shaded area shows the bulk states. The region shaded solid pink has a flat band of one dimensional edge modes, but is fully gapped in the bulk.} 
\label{spectrae}
\end{figure}

In Fig.\ \ref{fig:phasee} the extrinsic topological phase diagram is shown, with the positions used for the quenches in the following sections marked. The phases are divided by the edge gap closing line $|n|+|m|=1/2$; and the bulk gap closing lines $|m|=1$ and $2n=1+m^2$.  In the phase with flat bands they exist along all edges, see appendix \ref{app:mod} for examples of the band structure. In Fig.~\ref{spectrae} we show the single particle spectra $\epsilon$ along two cuts through the phase diagram. At $n=0.25$ we see the edge gap closing between four and two corner modes, followed by the bulk gap closing to the topologically trivial phase. At $n=0.8$ one can see the gapless phase containing flat bands which gives way to the phase with two corner modes, followed by the edge gap closing to the topologically trivial phase. 

\section{The Loschmidt Echo and Return Rate}\label{sec:los}

Dynamical quantum phase transitions can be traced to zeroes which occur in the Loschmidt echo, which cause non-analyticities in the associated return rate~\cite{Heyl2013}. In the form we are interested in here the Loschmidt echo is the overlap between an initial and a time evolved state, and we will follow the usual quantum quench protocol. In such a case we prepare the system in the ground state, $|\Psi_0\rangle$, of a Hamiltonian $\Hm^0=\Hm_{m_0,n_0,n'_0}$, and then time evolve it with respect to a new Hamiltonian $\Hm^1=\Hm_{m_1,n_1,n'_1}$. The Loschmidt echo is then
\begin{equation}
L(t)=\langle\Psi_0|e^{-i\Hm^1t}|\Psi_0\rangle\,,
\end{equation}
and the Loschmidt amplitude is the absolute magnitude of the Loschmidt echo. In the thermodynamic limit this quantity is exponentially suppressed in the system size, and it is natural to define the so-called return rate as
\begin{equation}
    l(t)=-\frac{1}{N}\ln\left|L(t)\right|\,,
\end{equation}
where $N$ is the system size, here the number of sites in the lattice. This is analogous to a free energy for the ``partition function'' $L(t)$ and it has a well defined $N\to\infty$ limit $l_0(t)=\lim_{N\to\infty}l(t)$.

For a simple two band free fermion model an analytical expression can be straightforwardly derived for translationally invariant systems~\cite{Vajna2015}. Generalisations to multi-band systems can in some cases be found~\cite{Maslowski2020}, however typically fully analytical expressions are no longer possible. As we are interested in boundary contributions for which we need a finite open system, factorisation of the Loschmidt echo in momentum space in any case fails. Instead we can use an alternative formation. Defining the correlation matrix as $\C_{ij}=\langle\Psi_0|\Psi^\dagger_i\Psi_j|\Psi_0\rangle$, where $i$ and $j$ run over a complete basis, then the Loschmidt echo is given by \cite{Levitov1996,Klich2003,Rossini2007}
\begin{equation}\label{rle}
L(t)=\det\underbrace{\left[1-\mathbf{\C}+\mathbf{\C}e^{i {\bm H}^1 t}\right]}_{\equiv {\bm M}(t)}\,.
\end{equation}
${\bm H}^1$ is the Hamiltonian matrix written in the same basis as $\C$, and we refer to ${\bm M}(t)$ as the Loschmidt matrix. When momentum is a good quantum number this trivially factorizes into momentum sub-spaces and one recovers the previously derived formulae. This makes it a convenient starting place for both the open and periodic systems when considering more than two bands, for which direct calculation of the overlap anyway becomes cumbersome. Our results here are based on this formulation, for bulk results we use momentum space where factorization into different momenta allows us to reach large system sizes, and for the thermodynamic limit to consider integrals over momentum.

It is often helpful to consider the eigenvalues of the Loschmidt matrix $\lambda_i(t)$ in terms of which we can write
\begin{equation}\label{ele}
L(t)=\prod_i\lambda_i(t)\,,
\end{equation}
and
\begin{equation}\label{rre}
l(t)=-\frac{1}{N}\sum_i\ln\left|\lambda_i(t)\right|\,.
\end{equation}
The non-analyticities in the return rate are determined by the zeros of the Loschmidt echo~\cite{Heyl2013} which occur when $L(t)=0$. These correspond to eigenvalues which become zero at critical times. Therefore one can analyze DQPTs directly from the behaviour of $\lambda_i(t)$. In one dimension the condition that $L(t_c)=0$ is satisfied by for a critical $i$ at a critical time. The non-analyticities only truly appear in the thermodynamic limit so let us now turn to the the bulk case. The critical $i$ corresponds to a critical momentum $k^*$, which along with $t_c$ will be the solution to the equation $\lambda_k^*(t_c)=0$. As $\lambda$ are complex this gives two equations for two unknowns.

In two dimensions the situation is different, as this equation can now be solved by a line of critical momenta ${\bm k}^*$. This results in an extended line of critical times. Therefore the DQPTs do not show up so clearly in $l(t)$ and one should consider its derivative~\cite{Vajna2015}. In terms of the eigenvalues one finds
\begin{equation}\label{retderdef}
d(t)\equiv\dot{l}(t)=-\frac{1}{N}\sum_{\bm k}\left|\frac{\dot{\lambda}(t)}{\lambda(t)}\right|\,,
\end{equation}
and for the Loschmidt matrix this becomes
\begin{equation}\label{retder}
d(t)\equiv\dot{l}(t)=-\frac{1}{N}\,\mathrm{Re}\left(\tr\left[\dot{\bm M}(t){\bm M}^{-1}(t)\right]\right)\,.
\end{equation}
In the following we will use $l(t)$, $d(t)$, and $\lambda_i(t)$ to investigate the DQPTs and the dynamical bulk boundary correspondence. For the thermodynamic limit we will use the convention $l_0(t)=\lim_{N\to\infty}l(t)$ and $d_0(t)=\lim_{N\to\infty}d(t)$, both of which can be calculated as two dimensional integrals over the momenta.

The zeroes of the Loschmidt echo can be understood as those Fisher zeroes in the complex plane which cross the real axis~\cite{Fisher1965,Heyl2013}. Generalizing to the complex $z$ plane we have
\begin{equation}
L(z)=\langle\Psi_0|e^{-i\Hm^1z}|\Psi_0\rangle\,,
\end{equation}
which give back the Loschmidt echo for $t=Re\,[z]$. As we can not solve $L(z)$ we use a proxy. Let $\lambda_0(z)$ be the eigenvalue with smallest magnitude. Then $L(z)=0$ if and only if $\lambda_0(z)=0$, and we can study $\lambda_0(z)$ numerically at a certain system size.

We will consider the following quench scenarios. We will use a convention where $A_z$ is a point in phase space with $A$ being either $I$ for intrinsic or $E$ for extrinsic and $z$ the number of corner modes. Here we will not quench into the flat band regions of the phase diagram where higher order topology is not the relevant ordering principle. For the specific parameters used see Table \ref{tab:quenches}, and see Figs.~\ref{fig:phasei} and \ref{fig:phasee} for their locations in the phase diagram. We consider all quenches between $I_4$, $I_2$, and $I_0$ and all quenches between $E_4$, $E_2$, and $E_0$. We also tested quenches $I_z\leftrightarrow E_z$ for $z\in\{0,2,4\}$. We note that our model is simple enough that we have only found DQPTs when we quench between different topological phases.

\begin{table}
    \begin{center}
    \begin{tabular}{|c|c|c|c|}
        \hline Label & Parameters & Corner modes & Type\\
        $A_z$ & $(m,n,n')$ & $z$ &  \\\hline
        $I_4$ & (0.4,0.1,0) & 4 & Intrinsic \\\hline
        $I_2$ & (0.4,0.4,0) & 2 & Intrinsic \\\hline
        $I_0$ & (1.2,0.1,0) & 0 & Intrinsic \\\hline
        $E_4$ & (0.4,0.1,0.1) & 4 & Extrinsic \\\hline
        $E_2$ & (0.4,0.4,0.4) & 2 & Extrinsic \\\hline
        $E_0$ & (1.2,0.1,0.1) & 0 & Extrinsic \\\hline
    \end{tabular}\caption{The points in the phase diagrams used for the quenches, see Figs.\ \ref{fig:phasei} and \ref{fig:phasee}. along with the number of corner modes they possess. As only one pint in each topological phase is considered in this article, each point in parameter space is given a simple label for convenience.}
    \label{tab:quenches}
    \end{center}
\end{table}

\section{Dynamical Quantum Phase Transitions}\label{sec:dqpts}

To find the DQPTs we start by considering the Fisher zeroes. In this section we will focus on results for the extrinsic case. For the intrinsic case we see similar results, and we present some in Appendix \ref{app:dqpt}. In Fig.~\ref{fig:fishex} we show the magnitude of the lowest eigenvalue of the Loschmidt matrix on a log scale for $z$. For $L(z)=0$ this should diverge, but at finite system sizes will just be large. Four examples are plotted. For the quench $E_2\to E_4$ there are no DQPTs, whereas for the other cases plotted DQPTs are present. A full list of when DQPTs occur, and whether they are periodic or aperiodic is give in Table \ref{tab:dqpts}. In those cases where there are DQPTs it is less clear whether they can be removed by continuously deforming the positions. For the quenches $E_4\leftrightarrow E_0$ it may be that the disappearance of the zeroes for large $Im[z]$ is a finite size effect. For $E_0\to E_2$ it seems that the Fisher zeroes cover only a finite region of the $z$-plane. Results for longer times which show the periodic reappearance of the zeroes can be seen in Fig.~\ref{fig:fishlong} in Appendix \ref{app:dqpt}.

\begin{figure}
\includegraphics[width=0.49\columnwidth]{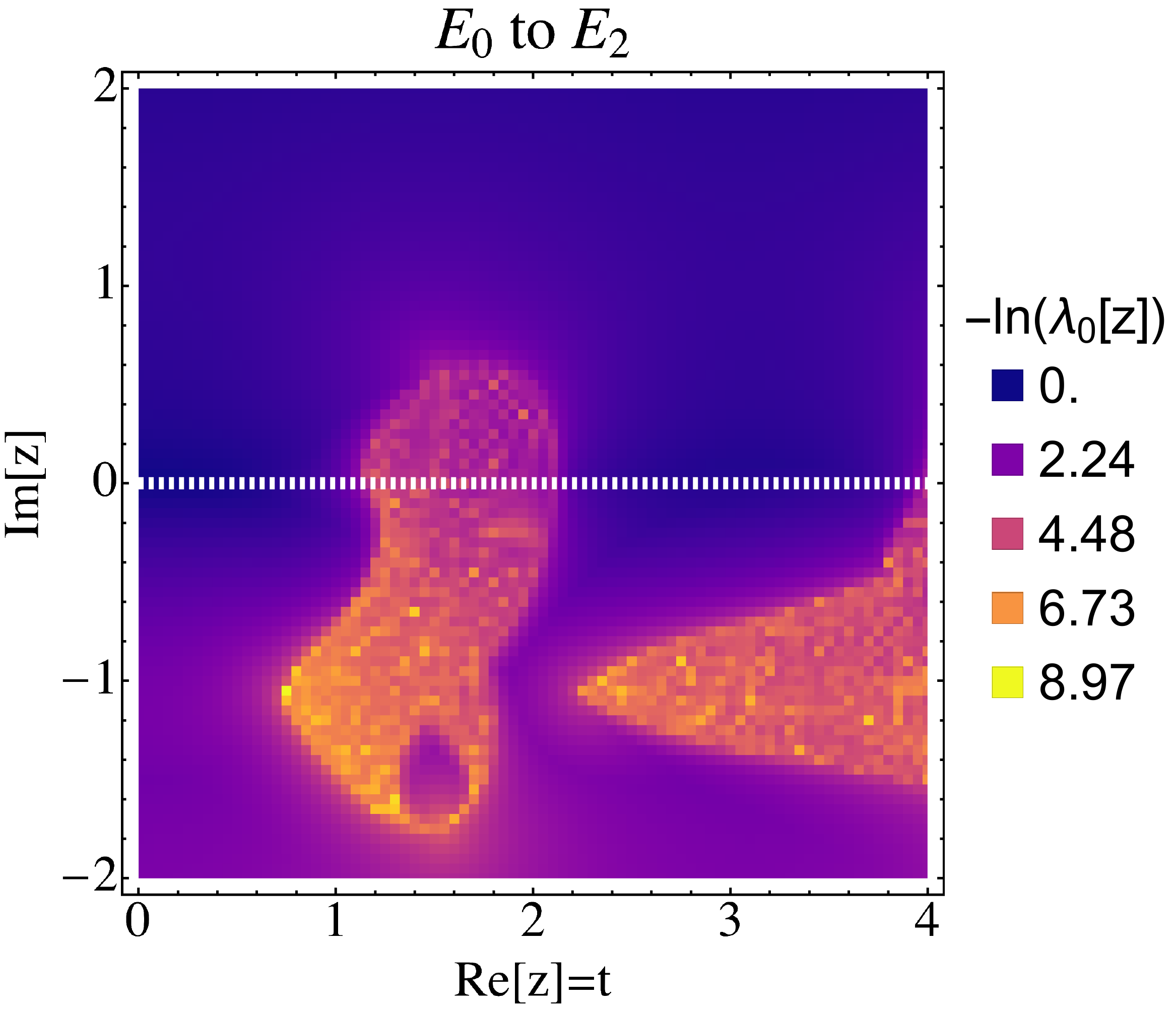}
\includegraphics[width=0.49\columnwidth]{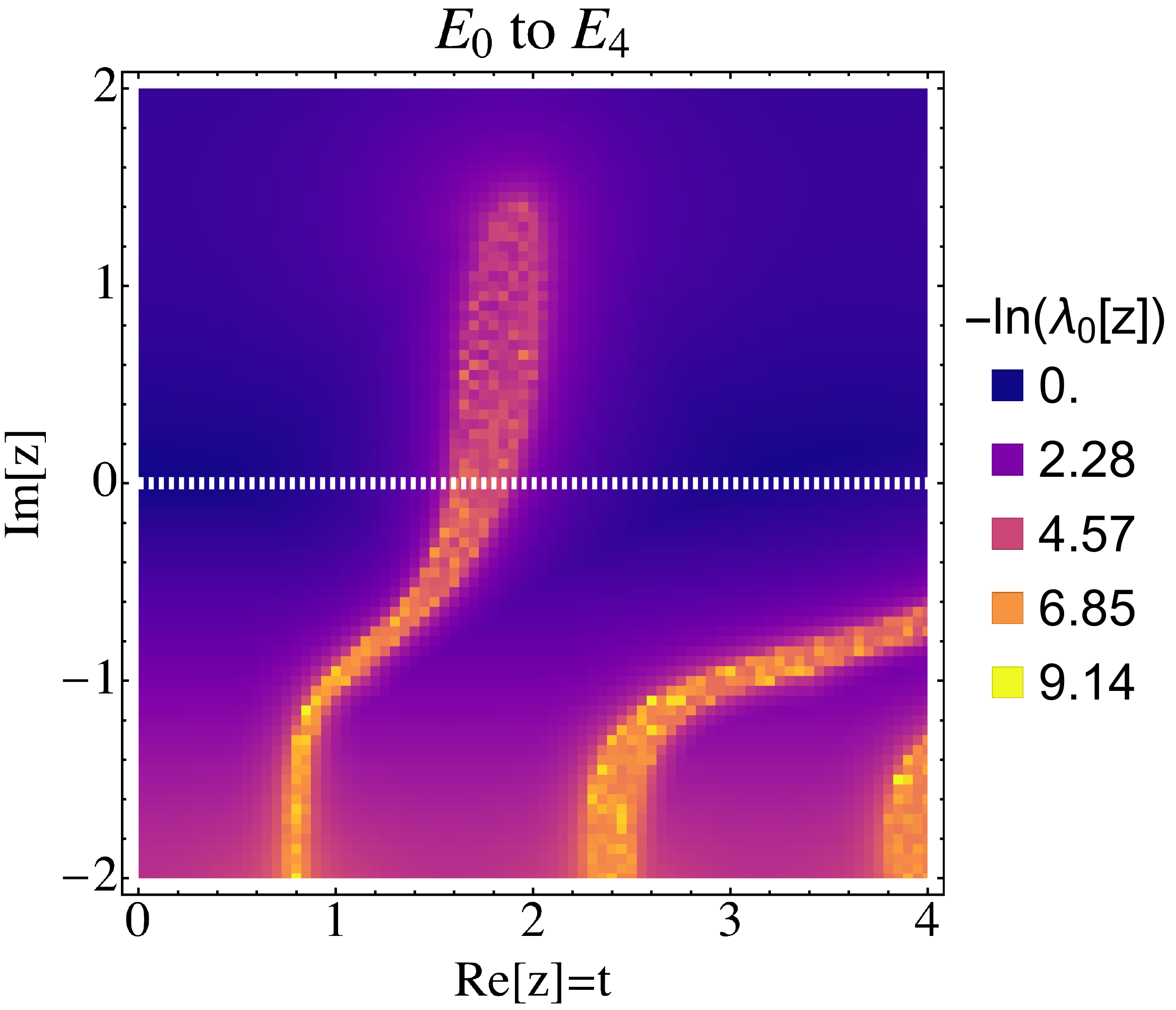}\\
\includegraphics[width=0.49\columnwidth]{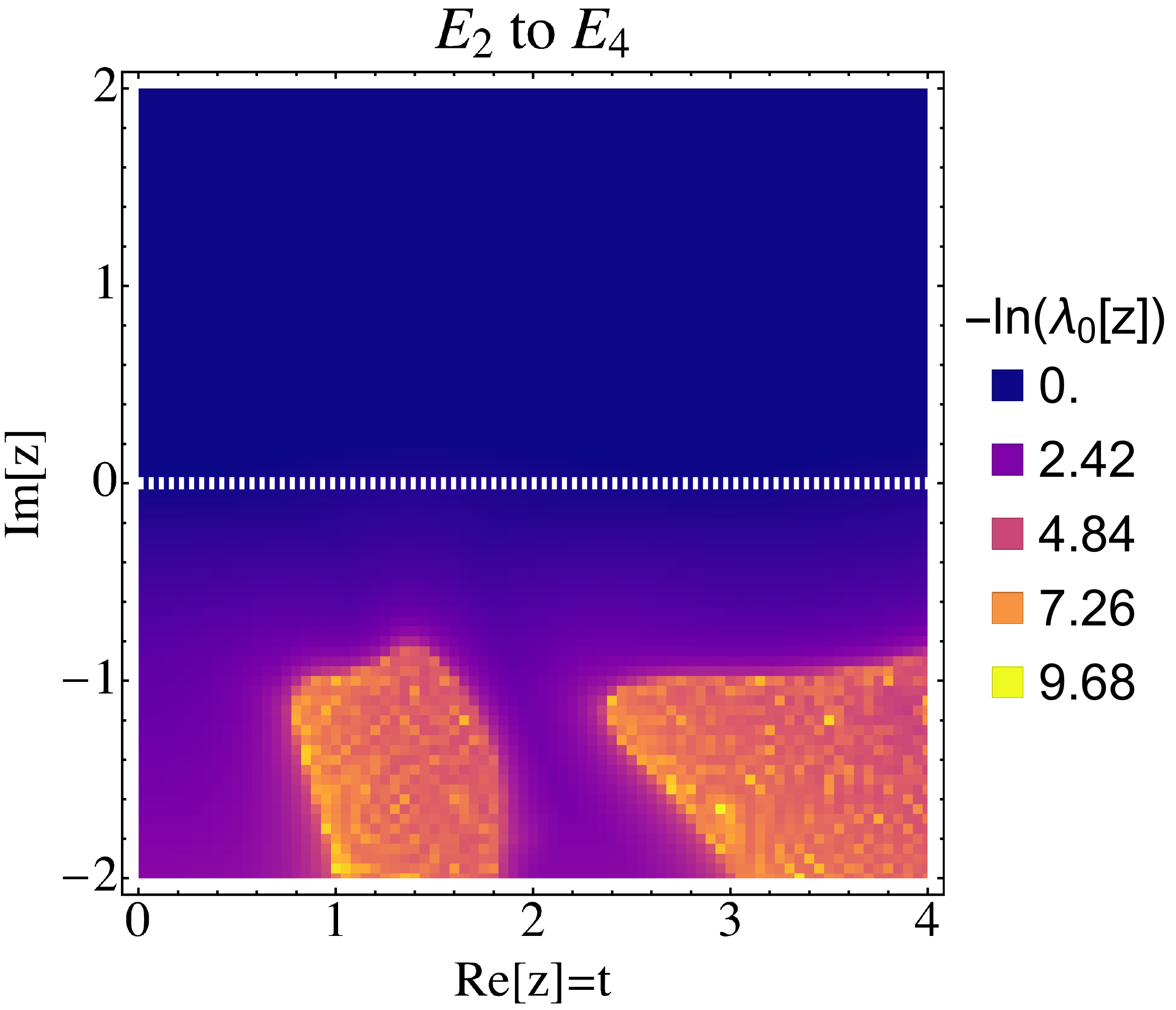}
\includegraphics[width=0.49\columnwidth]{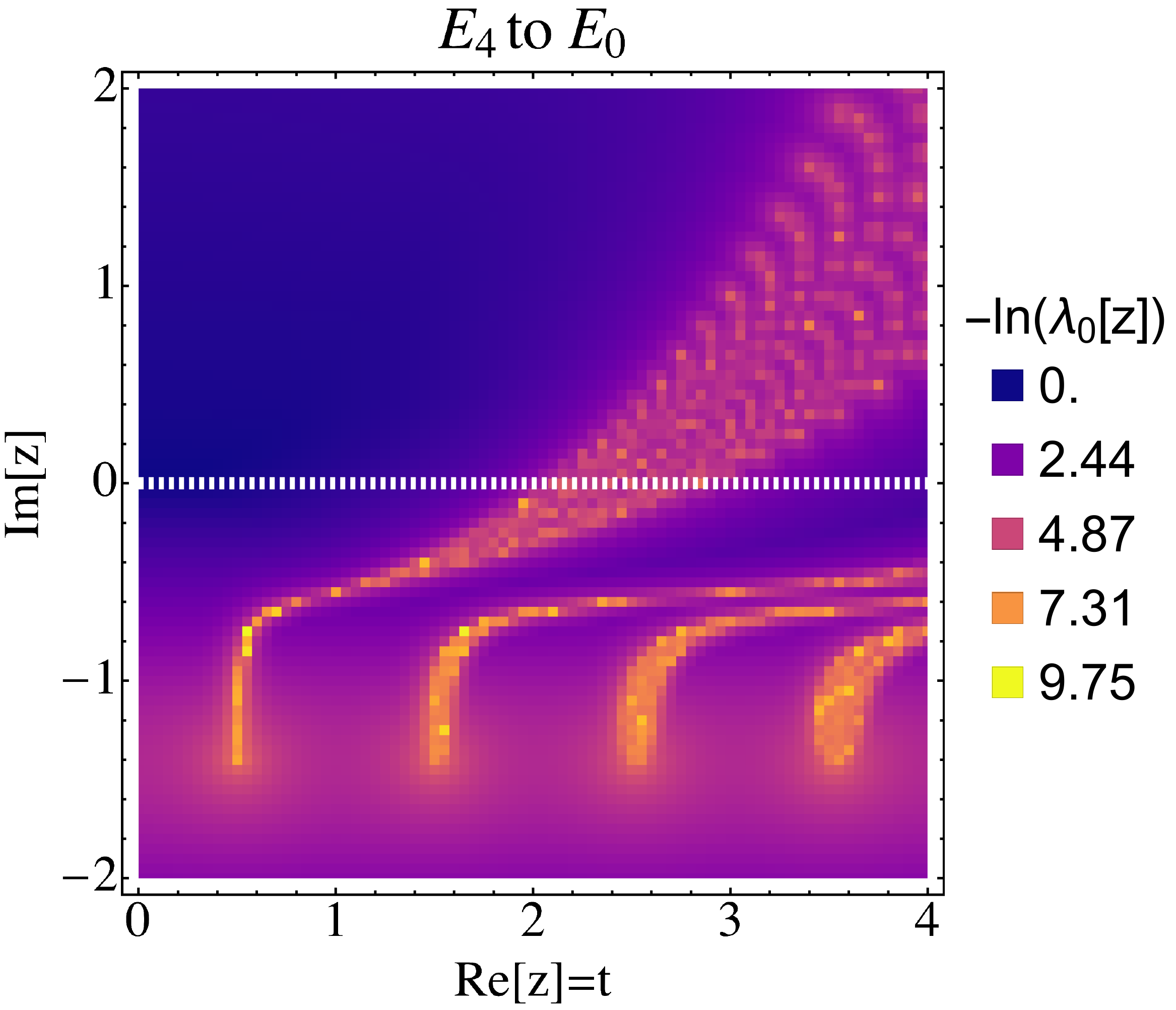}
\caption{Here we plot the proxy for the Fisher zeroes for four quenches, as labelled on the panels. Calculated for a system of size $N=202\times202$. In the thermodynamic limit Fisher zeroes would correspond to $-\ln[\lambda_0(z)]\to\infty$. For $E_2\to E_4$ the absence of DQPTs is clear, in all other cases DQPTs are present.}
\label{fig:fishex}
\end{figure}

\begin{figure*}
\includegraphics[width=0.95\columnwidth]{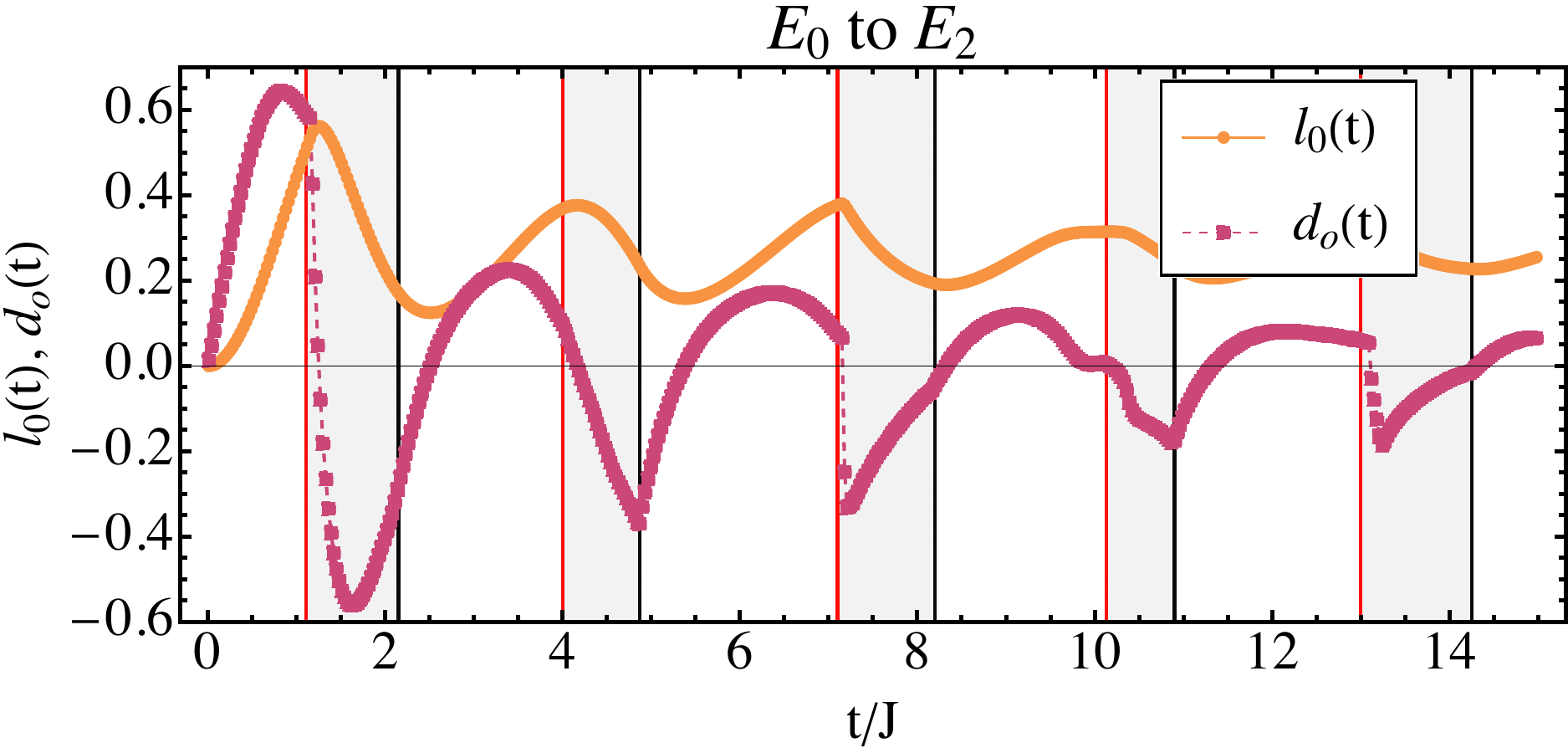}
\includegraphics[width=0.95\columnwidth]{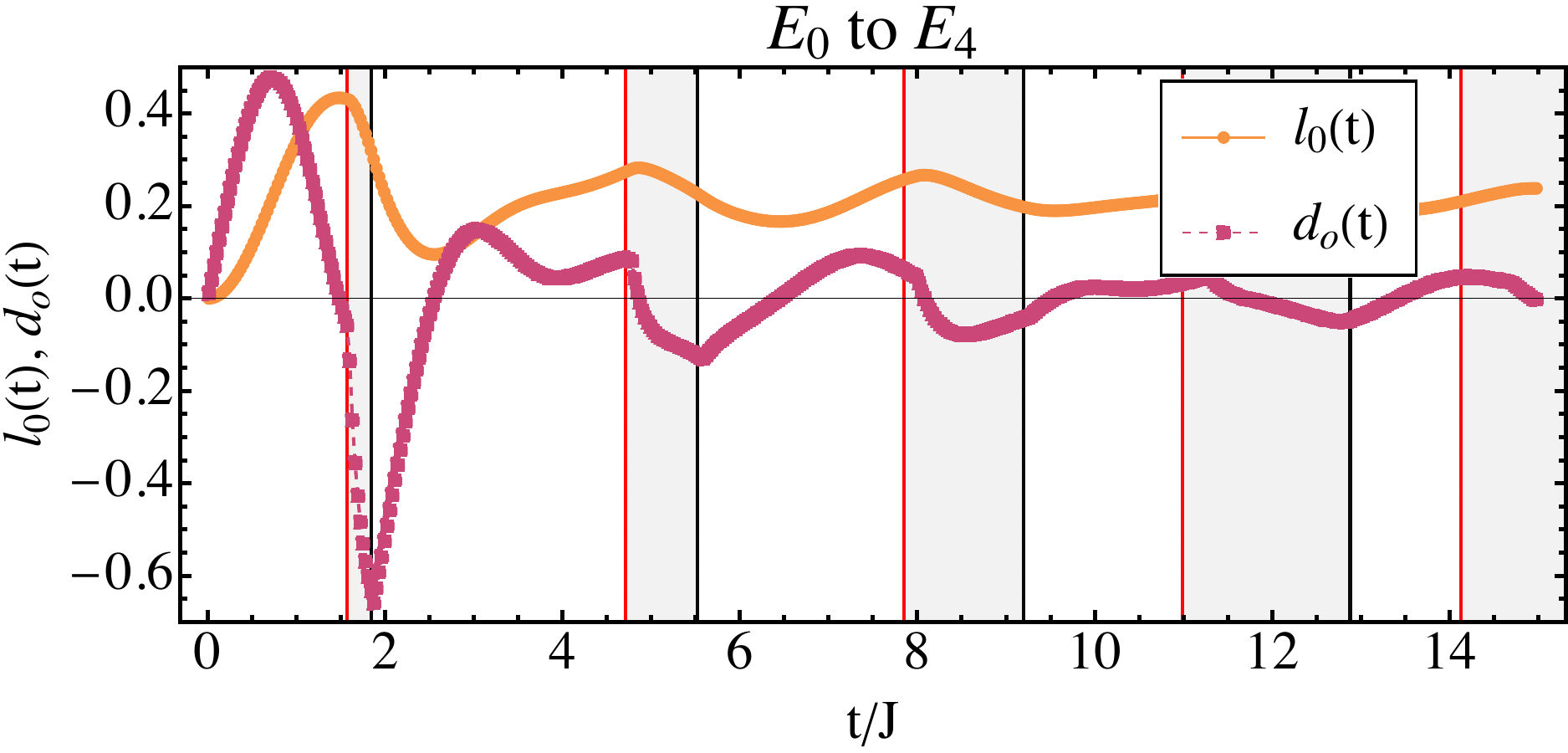}\\
\includegraphics[width=0.95\columnwidth]{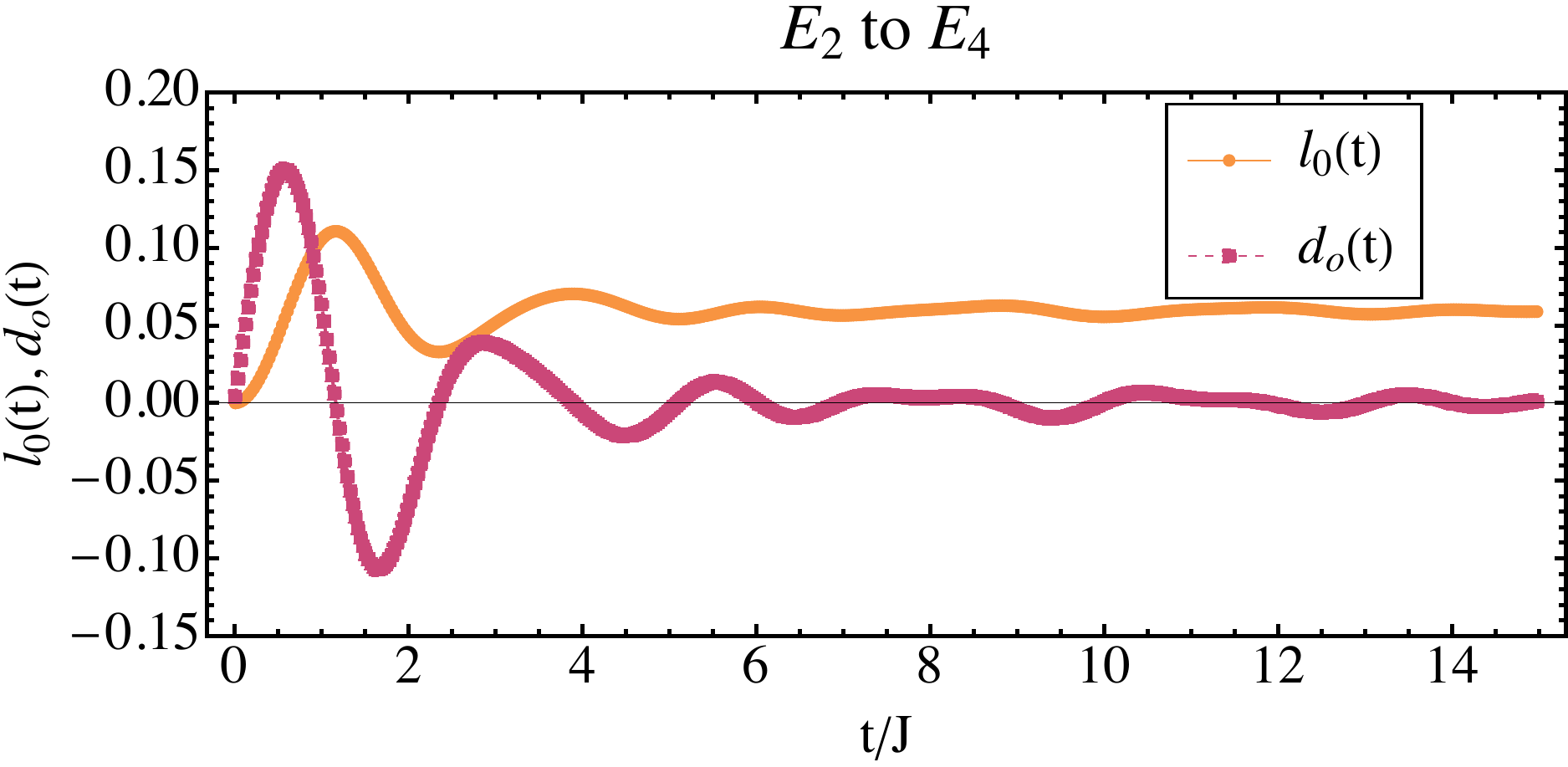}
\includegraphics[width=0.95\columnwidth]{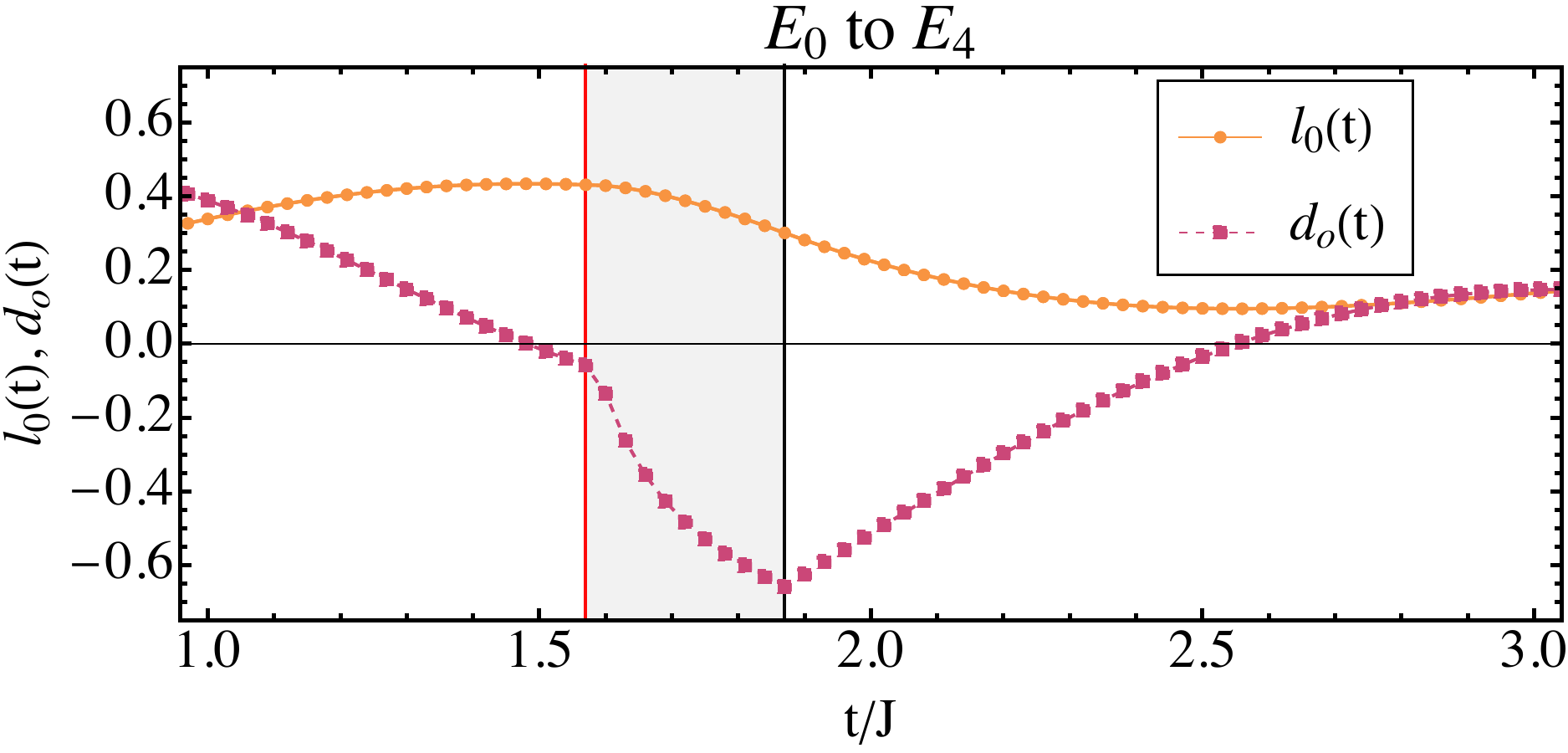}
\caption{The return rate $l_0(t)$ and its derivative $d_0(t)$ for several quenches of the extrinsic HOTI. The critical regions are shown in gray with $t_{c1}$ a red line and $t_{c2}$ a black line, where discontinuities in $d_0(t)$ are visible. This can be seen more clearly in the zoom-in of the quench $E_0\to E_4$. These critical times are calculated from the eigenvalues of $M(t)$ at a system size of $N=202\times202$.}
\label{fig:brrex}
\end{figure*}

Let us now turn to the return rate and its derivative. For the results in the thermodynamic limit, $l_0(t)$ and $d_0(t)$, we take Eqs.\ \eqref{rre} and \eqref{retder} in momentum space and perform the momentum integrals numerically. Due to the extended critical times, cusps are no longer expected in $l_0(t)$, rather we should see discontinuities in $d_0(t)$. In Fig.\ \ref{fig:brrex} we show several examples. These are all taken from the extrinsic case, similar results are found for the intrinsic HOTI, see Appendix \ref{app:dqpt} for examples. For quenches between the intrinsic and extrinsic HOTIs with the same number of corner modes, we see no DQPTs, though we stress here we tested examples where no critical line was crossed. For $E_0\to E_{2,4}$ DQPTs are clearly visible. We also show a zoom of a DQPT for $E_0\to E_4$. As an example of the lack of DQPTs we show the quench $E_2\to E_4$, where both $l_0(t)$ and $d_0(t)$ can be seen to be smooth functions. For quenches within a single phase we see no evidence of DQPTs, though we can not rule this out conclusively~\cite{Maslowski2020}.

Let $t_{c1}$ and $t_{c2}$ be the smallest and largest critical times for the first DQPT. In the simplest case we expect~\cite{Vajna2015} that the critical regions are therefore
\begin{equation}
    T_c=\bigcup_{m=0}^{\infty}T_{cm}=\bigcup_{m=0}^{\infty}\left[(2m+1)t_{c1},(2m+1)t_{c2}\right]\,,
\end{equation}
and for $t\in T_c$ there are zeroes eigenvalues of $M(t)$. Clearly the length of any contiguous critical region also grows as $(2m+1)(t_{c2}-t_{c1})$ with $m=0,1,2,\ldots$. As such after some time the regions start to overlap and it becomes difficult to discern their start and end. Here we have chosen quenches which delay this problem as much as possible. In Fig.\ \ref{fig:brrex} the critical regions are shown in gray with $t_{c1}$ a red line and $t_{c2}$ a black line. These critical times are calculated from the minimum eigenvalue of $M(t)$, which we label $\lambda_0(t)$, for a periodic system size of $N=202\times202$. When $\lambda_0(t)\leq\lambda^*$ we assume the system is critical with $\lambda^*$ a cut-off due to the finite size of our system. In the limit $N\to\infty$ we could take the condition $\lambda_0(t)=\lambda^*$. Examples of the eigenvalue behaviour are given in the next section on the dynamical bulk-boundary correspondence. In some cases we find that the critical times are no longer periodic, in which case each critical region must be calculated independently, and we label them as $\left[t_{c1m},t_{c2m}\right]$ with $m=1,2,3,\ldots$.

\begin{table}
    \begin{center}
    \begin{tabular}{|c|c|c|c|}
        \hline \hspace{0.1cm}$\left|\psi_0\right\rangle$\hspace{0.1cm} & \hspace{0.1cm}$\Hm^1$\hspace{0.1cm} & Critical line & Nature of \\
        $A_z$ & $A_z$ & crossed & critical cusps \\ \hline
        $E_4$ & $E_0$ & Bulk & Periodic \\\hline
        $E_0$ & $E_4$ & Bulk & Periodic \\\hline
        $E_2$ & $E_0$ & Bulk & Aperiodic \\\hline
        $E_0$ & $E_2$ & Bulk & Aperiodic \\\hline
        $E_4$ & $E_2$ & Edge & 0 \\\hline
        $E_2$ & $E_4$ & Edge & 0 \\\hline
        &&&\\\hline
        $I_4$ & $I_0$ & Edge & Periodic \\\hline
        $I_0$ & $I_4$ & Edge & Periodic \\\hline
        $I_2$ & $I_0$ & Edge & Aperiodic \\\hline 
        $I_0$ & $I_2$ & Edge & Aperiodic \\\hline
        $I_4$ & $I_2$ & Edge & 0 \\\hline
        $I_2$ & $I_4$ & Edge & 0 \\\hline
        &&&\\\hline
        $E_z$ & $I_z$ & None & 0 \\\hline
        $I_z$ & $E_z$ & None & 0 \\\hline
    \end{tabular}\caption{A list of all DQPTs found for the quenches studied. Listed firstly are the phases of the initial state and time evolving Hamiltonian, see main text for notation. We then note the type of critical line crossed by the quench and whether there are DQPTs, $0$ means no DQPTs are seen, and if so whether they appear periodically in time. $z\in\{0,2,4\}$.}
    \label{tab:dqpts}
    \end{center}
\end{table}

\section{The Dynamical Bulk Boundary Correspondence}\label{sec:dbb}

As we are interested in the boundary contributions to the return rate in principle we must consider the first correction to the thermodynamic limit:
\begin{equation}\label{bbreturn}
l(t)\sim l_0(t)+\frac{l_{B}(t)}{N}\,,
\end{equation}
with $l_0(t)$ and $l_{B}(t)$ the bulk and boundary contributions respectively. In principle $l_{B}(t)$ can be found from a finite size scaling analysis~\cite{Sedlmayr2018}, though in practice this is not always feasible. Due to the limited system size it is possible to reach for the two dimensional systems studied here a finite size scaling analysis is unfeasible. We note that a contribution to this difficulty is the necessity for multi-point precision to correctly describe the zero modes, which severely limits the system sizes that can be reached with reasonable memory capabilities and calculation times. In such a case we can use the behaviour of the $\lambda_i(t)$ as a proxy~\cite{Sedlmayr2018,Maslowski2020}. The dynamical bulk boundary correspondence states that for DQPTs with $\Hm^1$ belonging to a topologically non-trivial phase, $l_B(t)$ will exhibit characteristic plateaus between critical times. These plateaus are caused by eigenvalues of $M(t)$ which become pinned to zero between the critical times. For one dimensional topological insulators and superconductors one can show~\cite{Sedlmayr2018,Maslowski2020} that taking these zero modes $\lambda_n(t)$, where $n=0,1,2,\ldots \tilde n-1$ one finds that
\begin{equation}\label{lapprox}
    l(t)-l_0(t)\approx -\frac{1}{N}\sum_{n=0}^{\tilde n-1}\ln\left|\lambda_n(t)\right|
\end{equation}
where $l(t)$ is calculated for a system size of $N$. $\tilde n$ is the number of modes which become pinned to zero.

Here we focus on the appearance, or not, of these zero modes. As the HOTI DQPT already results in extended times of $\lambda_n(t)\approx$ a direct comparison of $l(t)-l_0(t)$ and $\ln\left|\lambda_n(t)\right|$ at available system sizes is not possible, and we focus purely on the behaviour of the $\lambda_n(t)$. In the following we show results for both an open nanoflake of size $N=50\times 50$, which has the corner modes present in the appropriate phases, and a periodic `bulk' system of size $N=202\times 202$. For the bulk case we plot only the smallest eigenvalue $|\lambda_n(t)|$, and the lowest four for the open systems.

\begin{figure}
\includegraphics[width=0.95\columnwidth]{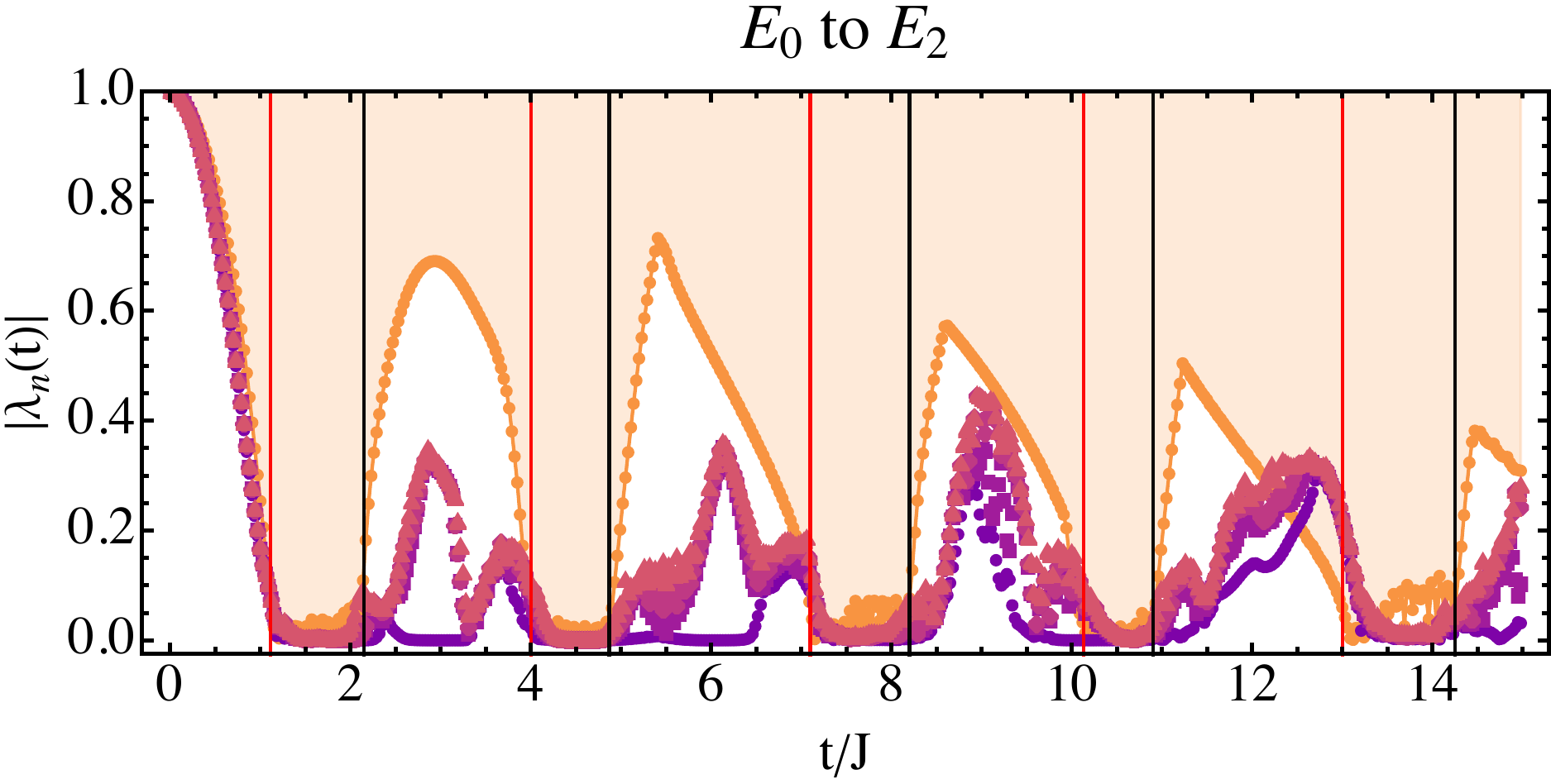}
\includegraphics[width=0.95\columnwidth]{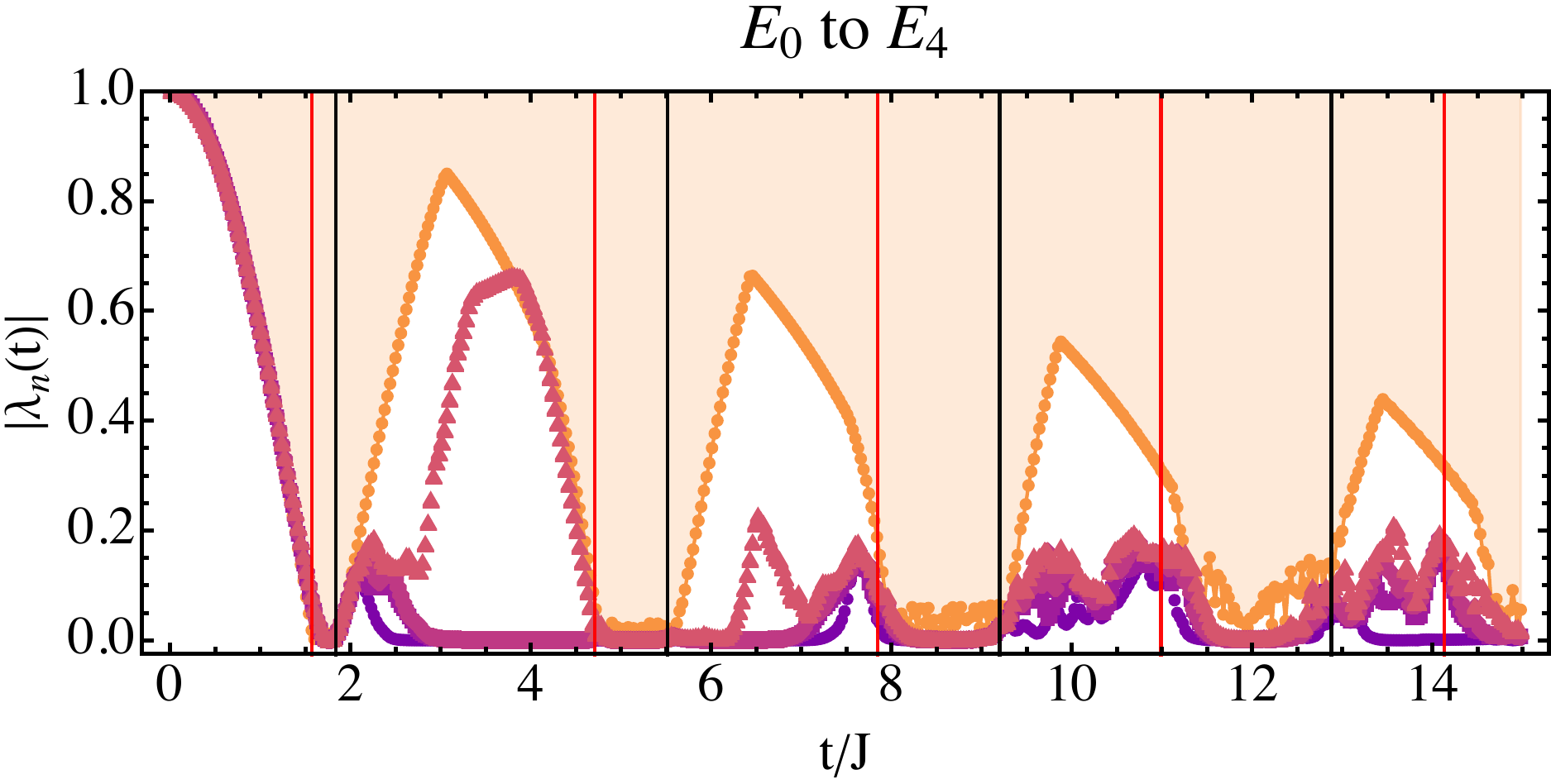}
\caption{Plots of the absolute values of the smallest eigenvalues of $M(t)$, $|\lambda_n(t)|$ for two quenches of the extrinsic HOTI, as labelled on the panels. Shown are the lowest four eigenvalues for an open nanoflake of size $N=50\times 50$ and the lowest value for a periodic system of size $N=202\times 202$, orange dots. The shaded region is the region where eigenvalues of $M(t)$ exist for the bulk system. In both of these cases DQPTs are present, see Fig\ \ref{fig:brrex}, and we expect pinned zero eigenvalues $\lambda_n(t)$ due to the dynamical bulk-boundary correspondence.}\label{fig:evex} 
\end{figure}

\begin{figure}
\includegraphics[width=0.95\columnwidth]{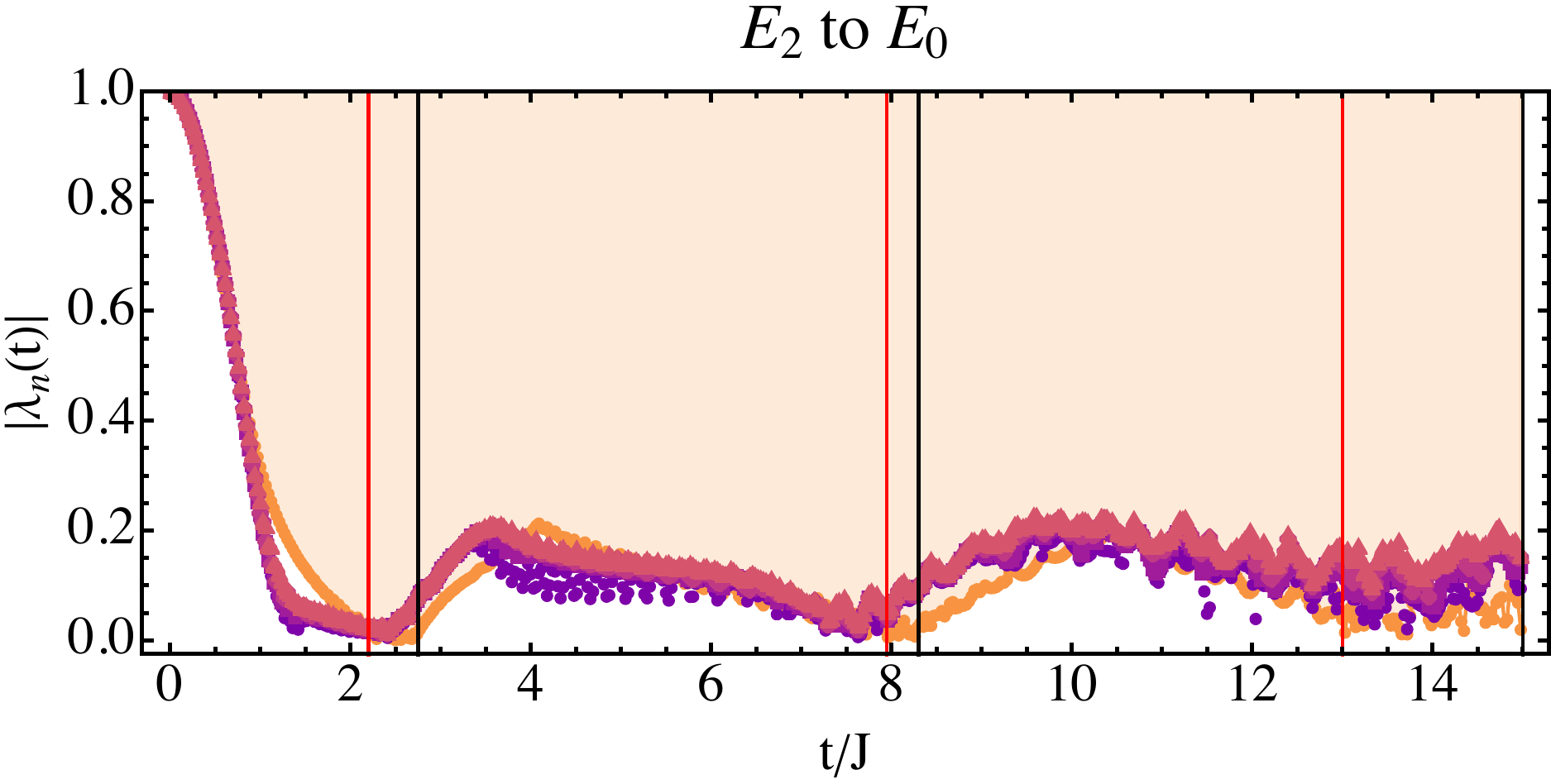}
\includegraphics[width=0.95\columnwidth]{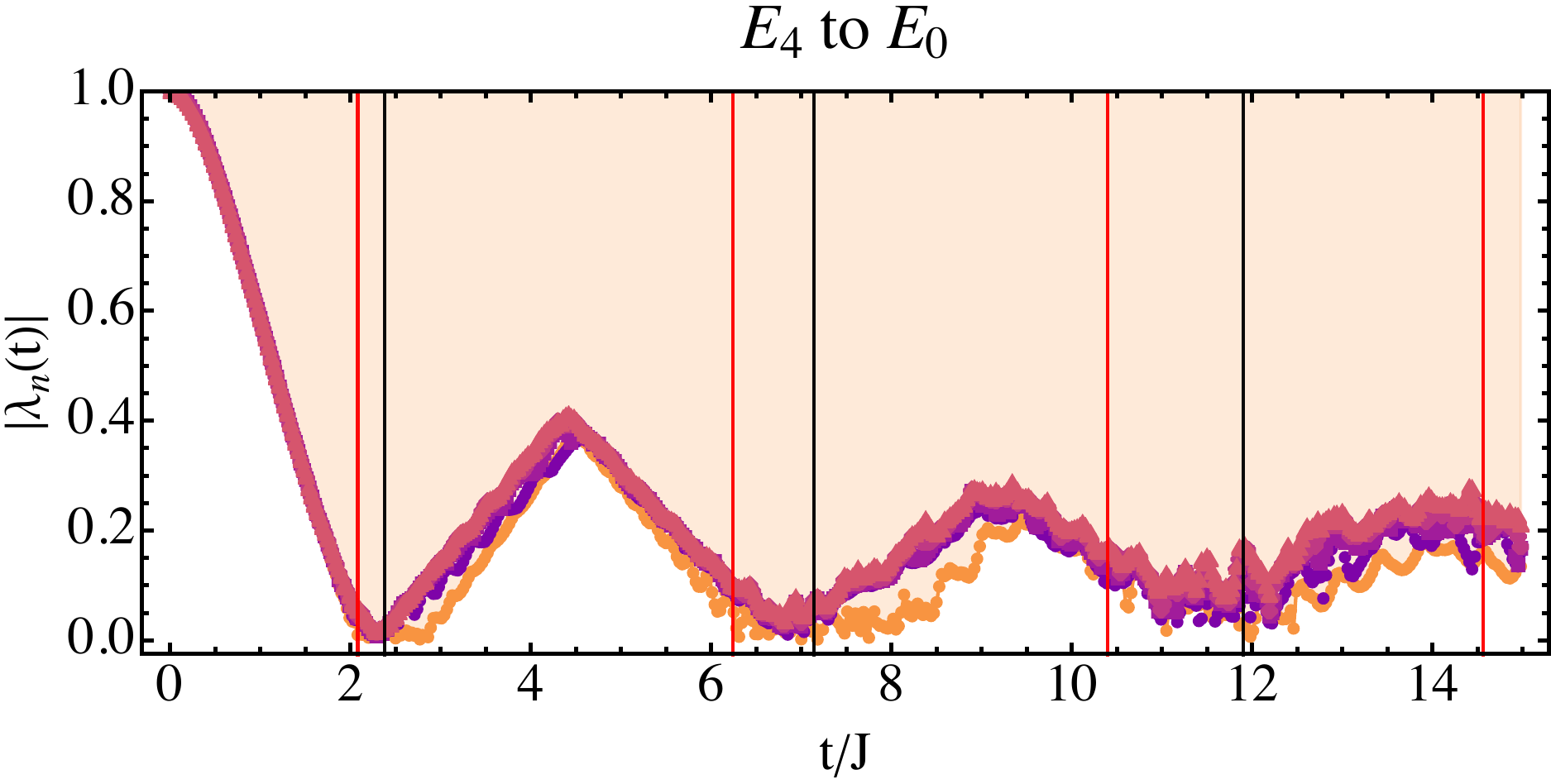}
\caption{Plots of the absolute values of the smallest eigenvalues of $M(t)$, $|\lambda_n(t)|$ for two quenches of the extrinsic HOTI, as labelled on the panels. Shown are the lowest four eigenvalues for an open nanoflake of size $N=50\times 50$ and the lowest value for a periodic system of size $N=202\times 202$, orange dots. The shaded region is the region where eigenvalues of $M(t)$ exist for the bulk system. As expected there are no pinned zero eigenvalues outside of $T_c$, in agreement with the dynamical bulk-boundary correspondence.}\label{fig:evex2} 
\end{figure}

In Fig.~\ref{fig:evex} the eigenvalues of the Loschmidt matrix are shown for two quenches into the topologically non-trivial phases where DQPTs are present for the extrinsic HOTI. In both cases zero eigenvalues can be seen. Firstly for the quench $E_0\to E_2$ (approximately) a single zero eigenvalue occurs between the critical regions $T_{c0}$ and $T_{c1}$ and also between $T_{c1}$ and $T_{c2}$. There are then no zeroes present $T_{c2}$ and $T_{c3}$ or between $T_{c3}$ and $T_{c4}$. After $T_{c4}$ they may reappear, but the data is already not very clear. For the quench $E_0\to E_4$ (approximately) three zero eigenvalues occur between the critical regions $T_{c0}$ and $T_{c1}$ and also between $T_{c1}$ and $T_{c2}$. After this it becomes hard to be confident on whether they exist or not. This tentative `double presence' then 'double absence' is already different from previous behaviour seen in one dimensional topological systems. In one dimension, for quenches to a topological phase with winding number 1, two zero eigenvalues appear periodically between critical times~\cite{Sedlmayr2018}. For larger winding numbers more zero eigenvalues are present, and the critical times at which they appear and disappear becomes more complicated~\cite{Sedlmayr2018,Maslowski2020}. We can also check that for the quenches $E_2\to E_0$ and $E_4\to E_0$ there are no pinned zero eigenvalues outside of $T_c$, see Fig.~\ref{fig:evex2}.

\begin{figure}
\includegraphics[width=0.95\columnwidth]{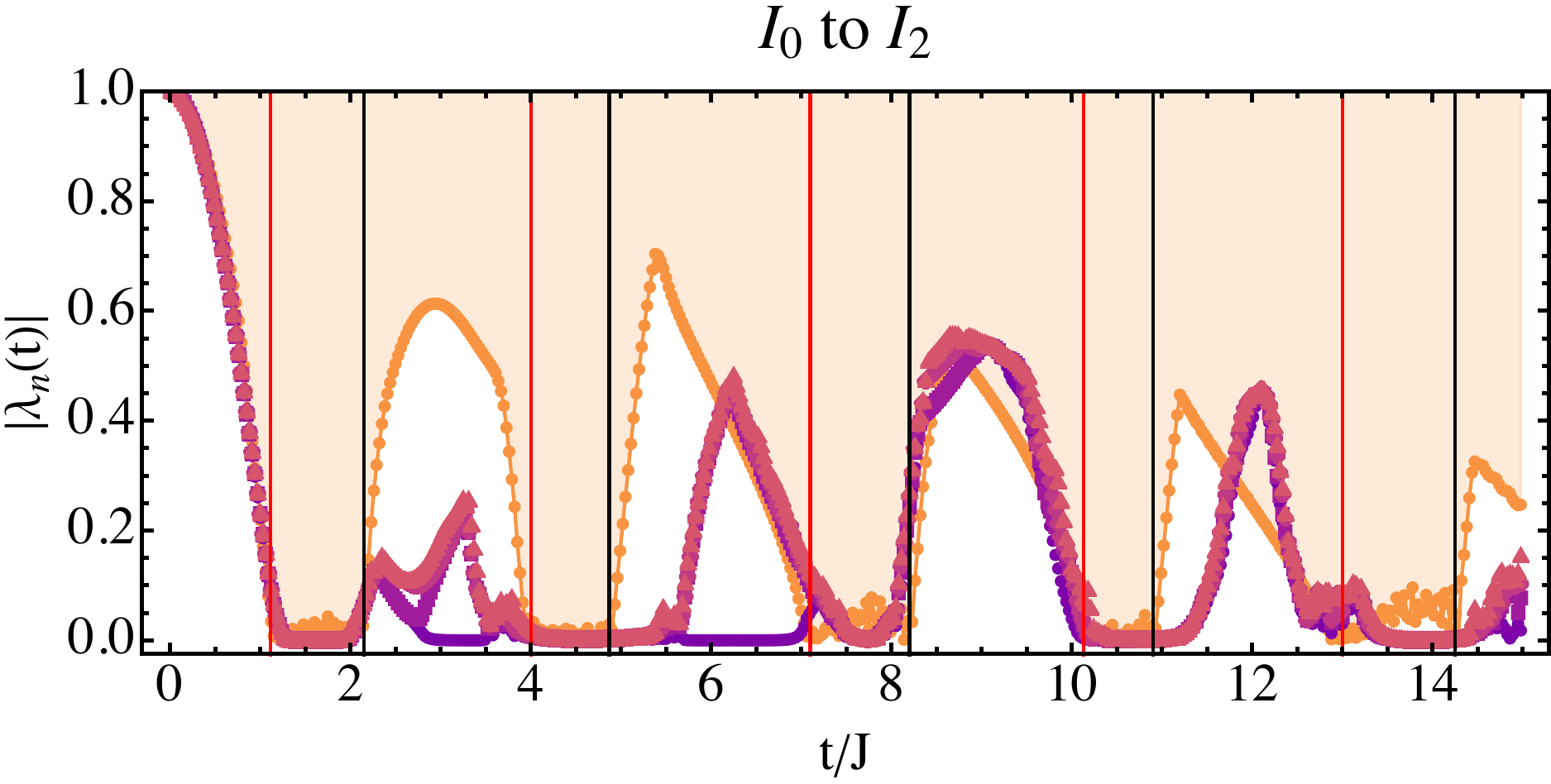}
\includegraphics[width=0.95\columnwidth]{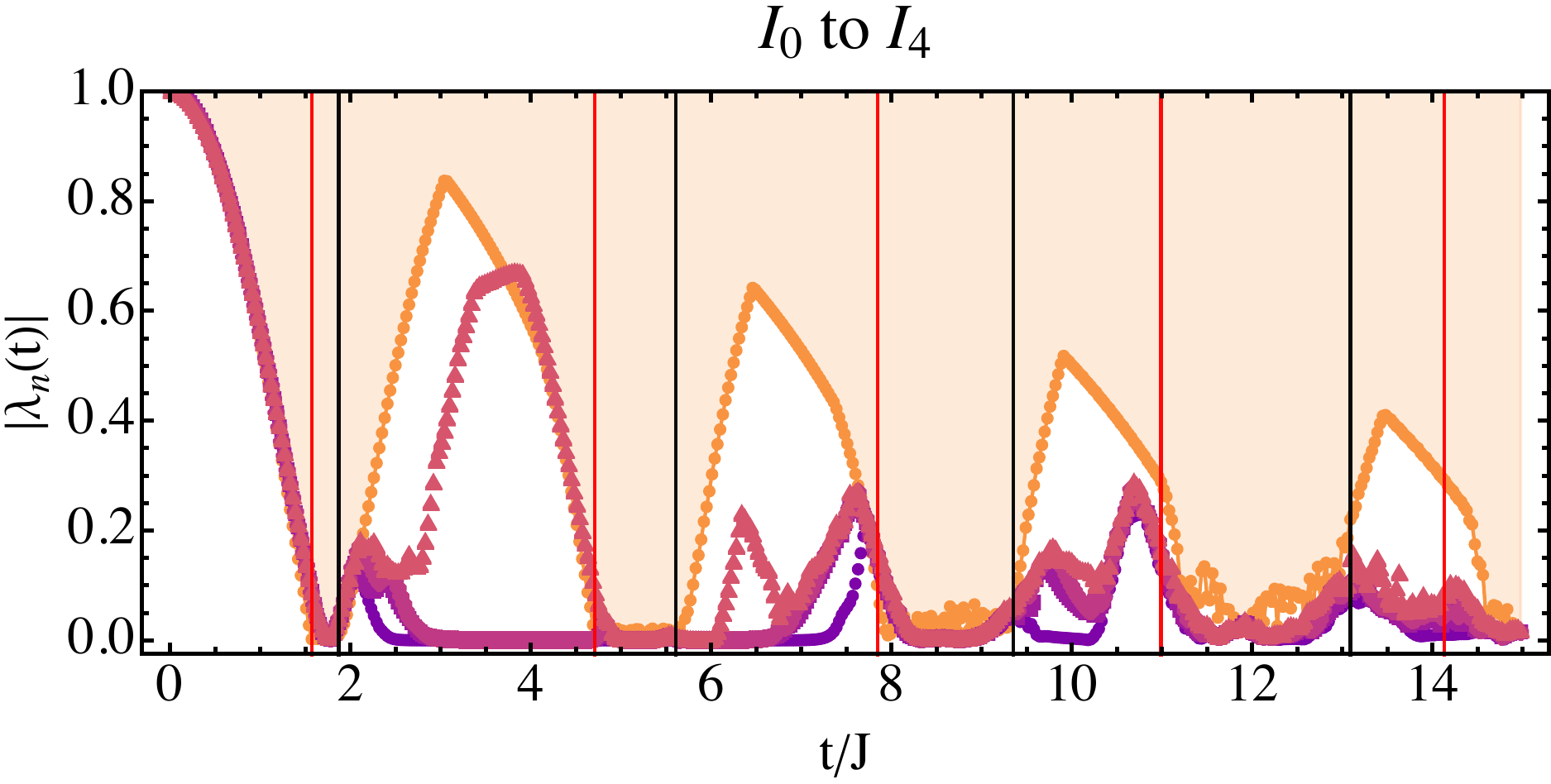}
\caption{Plots of the absolute values of the smallest eigenvalues of $M(t)$, $|\lambda_n(t)|$ for two quenches of the intrinsic HOTI, as labelled on the panels. Shown are the lowest four eigenvalues for an open nanoflake of size $N=50\times 50$ and the lowest value for a periodic system of size $N=202\times 202$, orange dots. The shaded region is the region where eigenvalues of $M(t)$ exist for the bulk system.}\label{fig:evin} 
\end{figure}

\begin{figure}
\includegraphics[width=0.95\columnwidth]{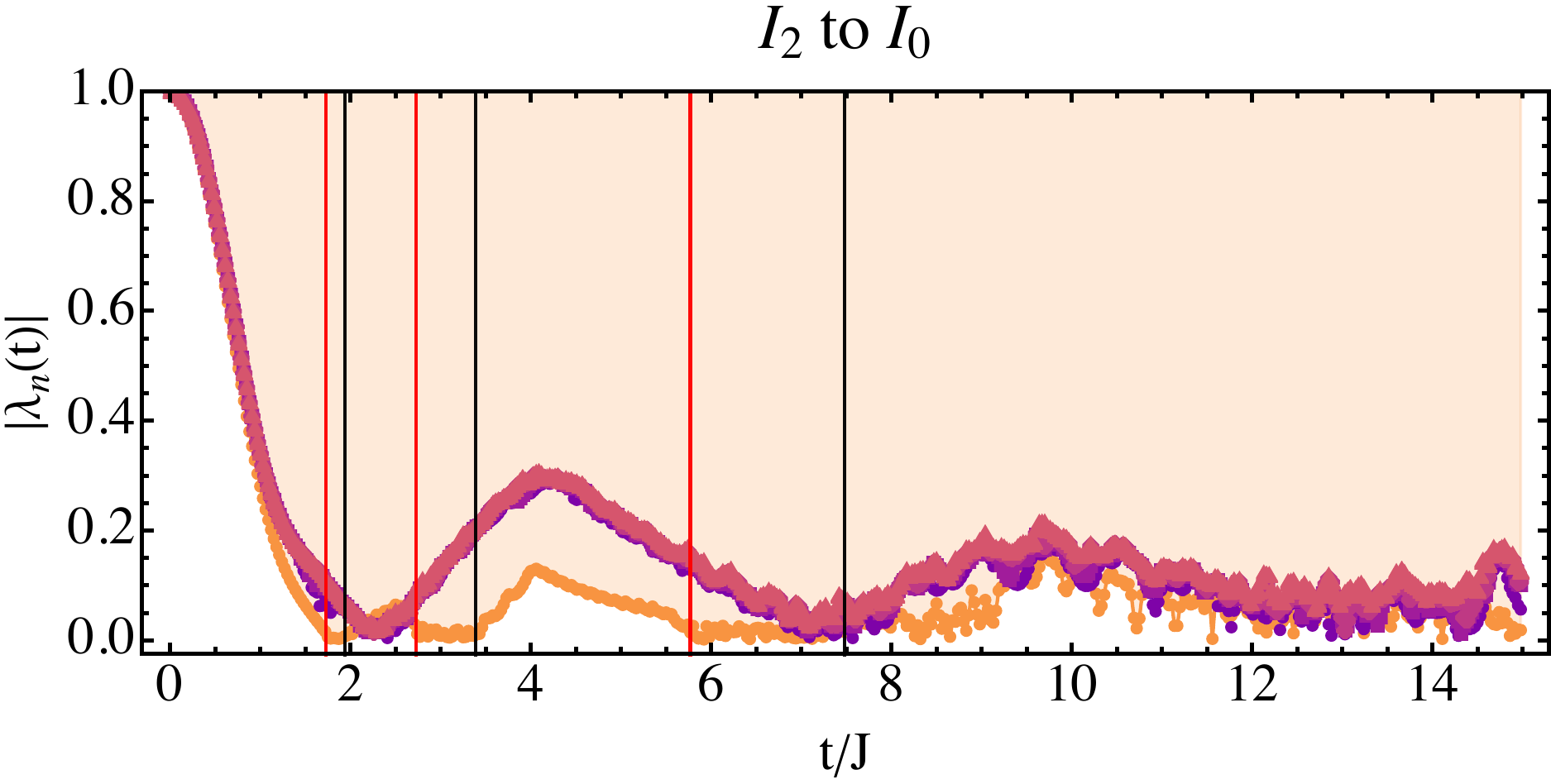}
\includegraphics[width=0.95\columnwidth]{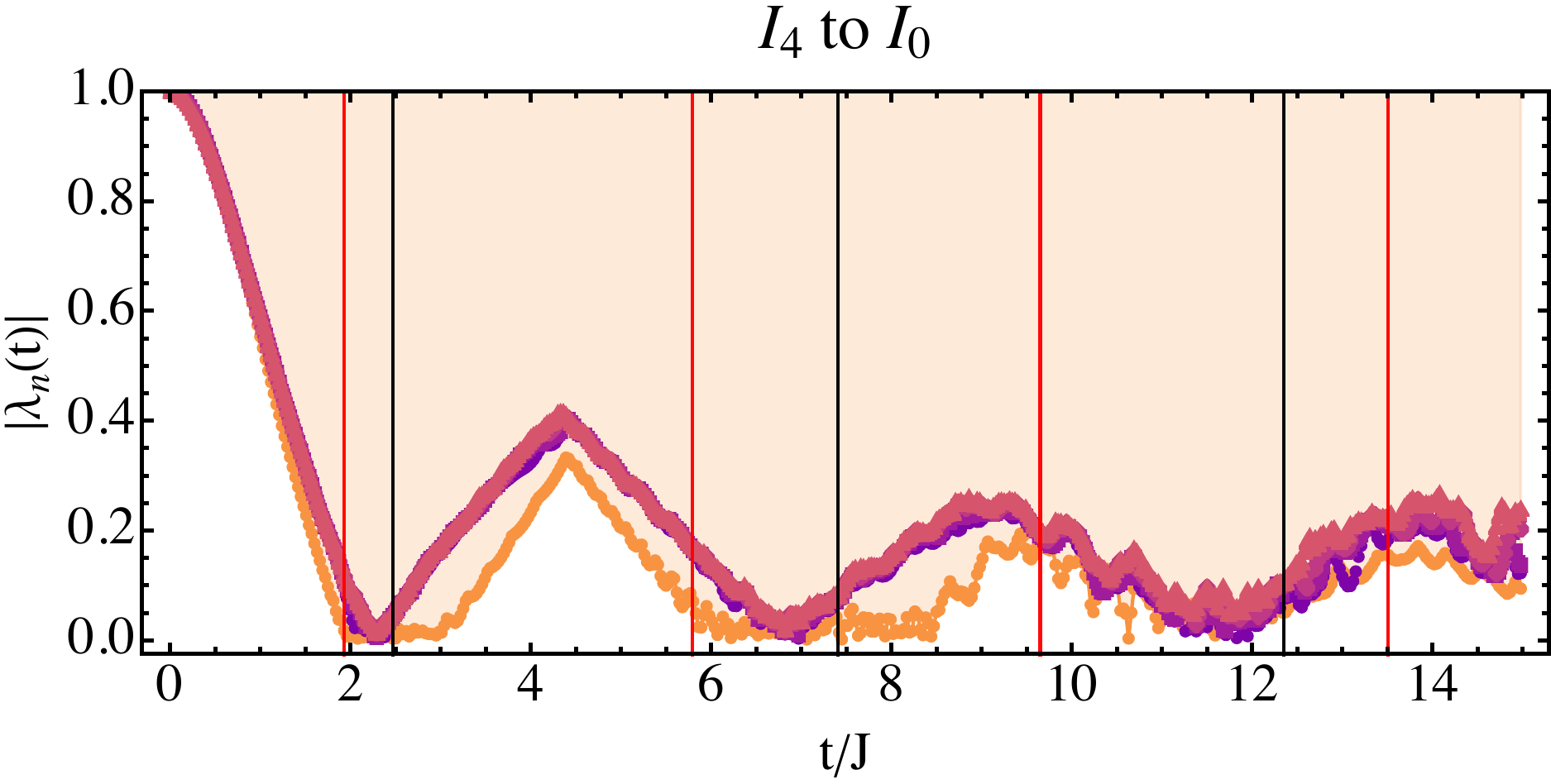}
\caption{Plots of the absolute values of the smallest eigenvalues of $M(t)$, $|\lambda_n(t)|$ for two quenches of the intrinsic HOTI, as labelled on the panels. Shown are the lowest four eigenvalues for an open nanoflake of size $N=50\times 50$ and the lowest value for a periodic system of size $N=202\times 202$, orange dots. The shaded region is the region where eigenvalues of $M(t)$ exist for the bulk system. As expected there are no pinned zero eigenvalues outside of $T_c$, in agreement with the dynamical bulk-boundary correspondence.}\label{fig:evin2} 
\end{figure}

For the intrinsic case all quenches cross critical lines where only the edge gap closes. In Fig.~\ref{fig:evin} the lowest Loschmidt eigenvalues for quenches $I_0\to I_{2,4}$ are plotted. They show the same pattern as for the extrinsic HOTI in Fig.~\ref{fig:evex}. In accordance with the dynamical bulk-boundary correspondence the reverse quenches $I_{2,4}\to I_0$ have, within finite size errors, no zero eigenvalues outside of $T_c$, see Fig.\ \ref{fig:evin2}.

\section{Discussion and Conclusions}\label{sec:dis}

In this article we have extended the definition of DQPTs to higher order topological matter, focusing on two dimensional HOTIs with different numbers of corner modes. A general model was developed which allows us to reach a multitude of different phases with a single model. As for usual two dimensional topological insulators the DQPTs can be observed in the time derivative of the return rate. For quenches between the topologically trivial and non-trivial phases we find DQPTs. The critical regions $T_{cm}$ are periodic for quenches involving the four corner mode phases, and aperiodic for quenches involving the two corner mode phases. All other quenches investigated result in no DQPTs. We tested both an intrinsic and extrinsic HOTI, with qualitatively similar results in both cases. The model described here also possesses flat bands of one dimensional edge modes, and what role they may play in the dynamics, as well as how generic the results seen here are for HOTIs, would be interesting questions for further studies.

To summarise we see that eigenvalues of $M(t)$ become pinned to zero between critical regions $T_{cm}$ between DQPTs for quenches into the topologically non-trivial phases. This constitutes the main generalisation of the dynamical bulk-boundary correspondence to two dimensional HOTIs. The exact number of the zeroes and the critical regions between which they appear seem ordered, but the exact nature of that ordering is not clear. The phases with two corner modes appear to result in $\tilde n=1$ and those with four corner modes appear to result in $\tilde n=3$, see Eq. (\ref{lapprox}). Furthermore the zero eigenvalues appear and disappear not between successive $T_{cm}$ but on an alternative pattern. Which of these observations are generic, and which particular to the model here would be an interesting extension of this work. Also of interest would be any potential proof of the dynamical bulk-boundary correspondence and extensions to usual two dimensional topological insulators and crystalline topological insulators.

\acknowledgments
This work was supported by the National Science Centre (NCN, Poland) under the grant 2019/35/B/ST3/03625. NS gratefully acknowledges helpful and clarifying discussions with Piet Brouwer about higher order topology.

\appendix
\counterwithin{figure}{section}

\section{Some More Details on the Model}\label{app:mod}

In this appendix we give some more details about the model used throughout this article. First for convenience here we list some of the commutation properties of the matrices involved in the symmetry operations on the model. One can show that 
\begin{equation}
    \{\Gamma_3,\mathcal{U}_y\}=0\textrm{ and }
    \{\Gamma_{5,6,8},\mathcal{U}_y\}=0
\end{equation}
anti-commute, and that
\begin{equation}
    [\Gamma_7,\mathcal{U}_y]=0\textrm{ and }[\Gamma_{1,2,4},\mathcal{U}_y]=0
\end{equation}
commute.

\begin{figure}[t!]
\includegraphics[width=0.49\columnwidth]{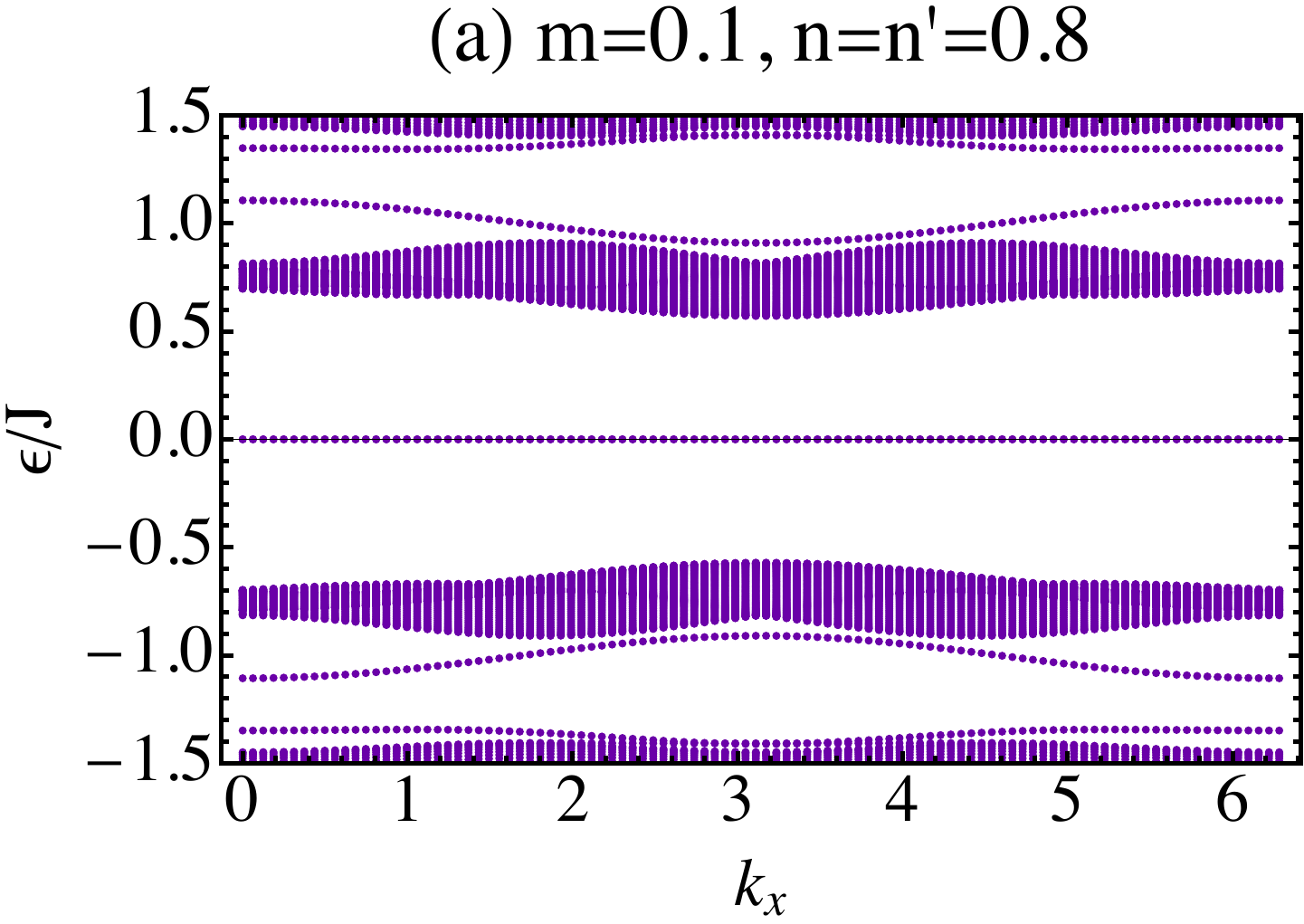}
\includegraphics[width=0.49\columnwidth]{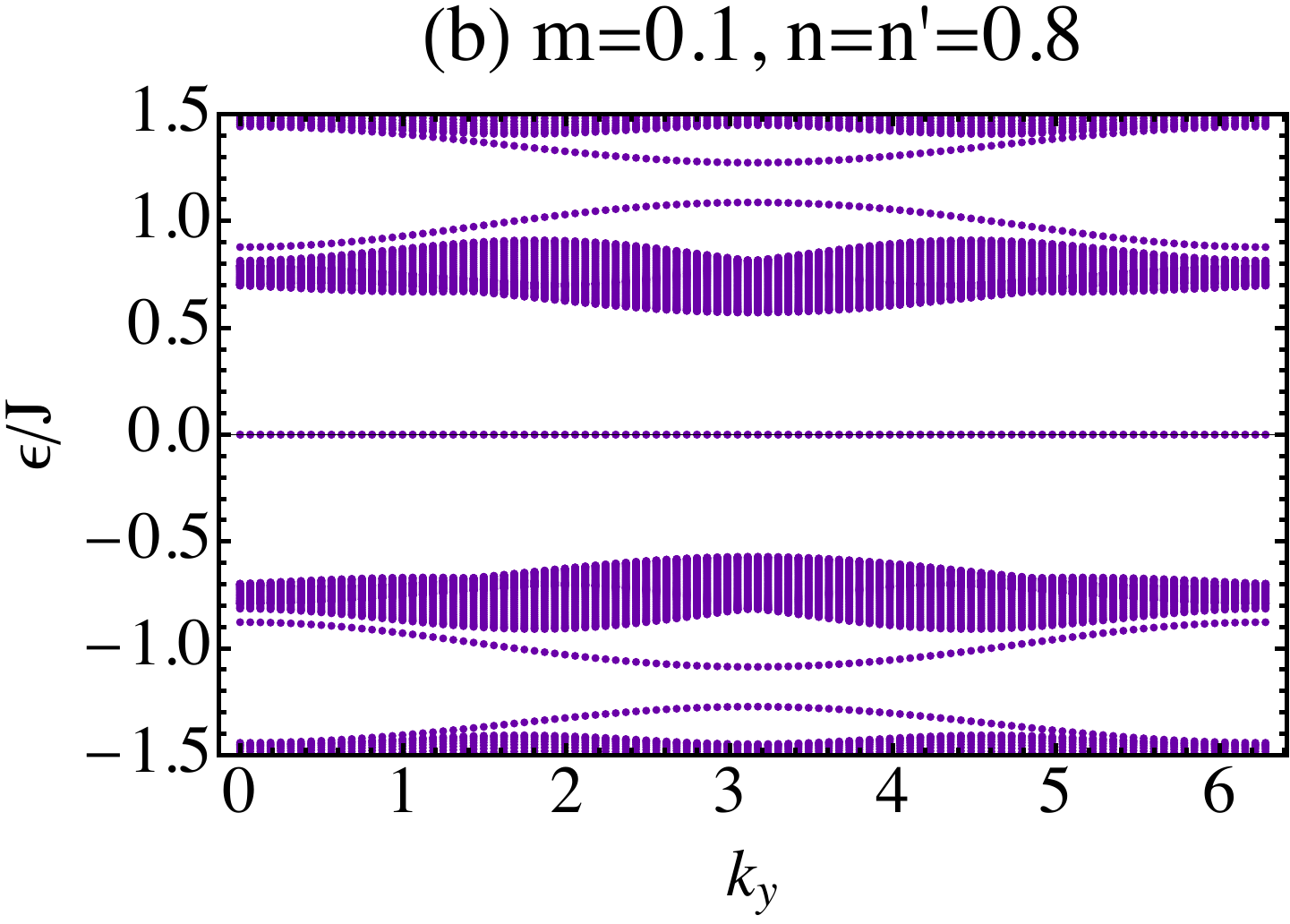}\\
\includegraphics[width=0.49\columnwidth]{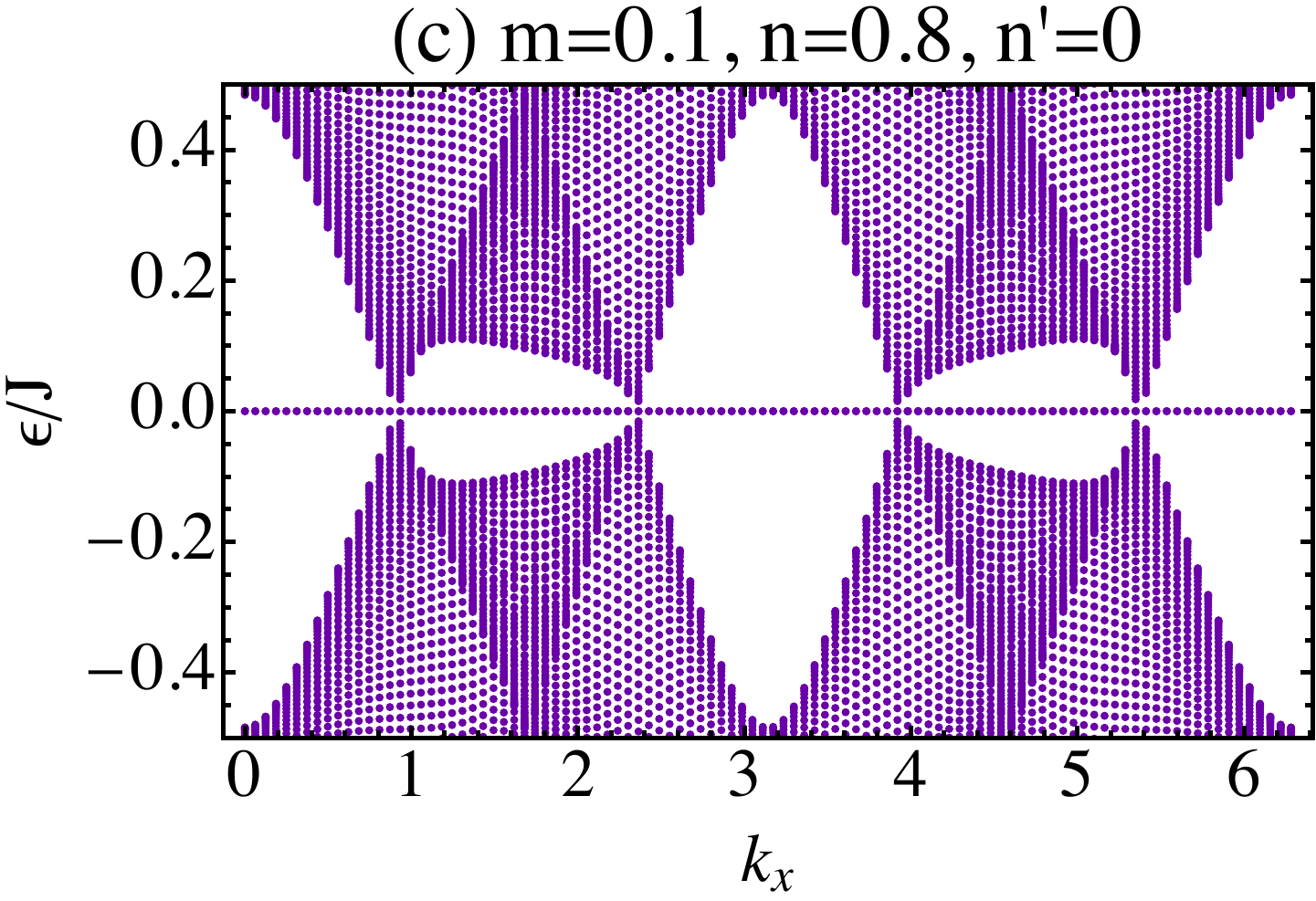}
\includegraphics[width=0.49\columnwidth]{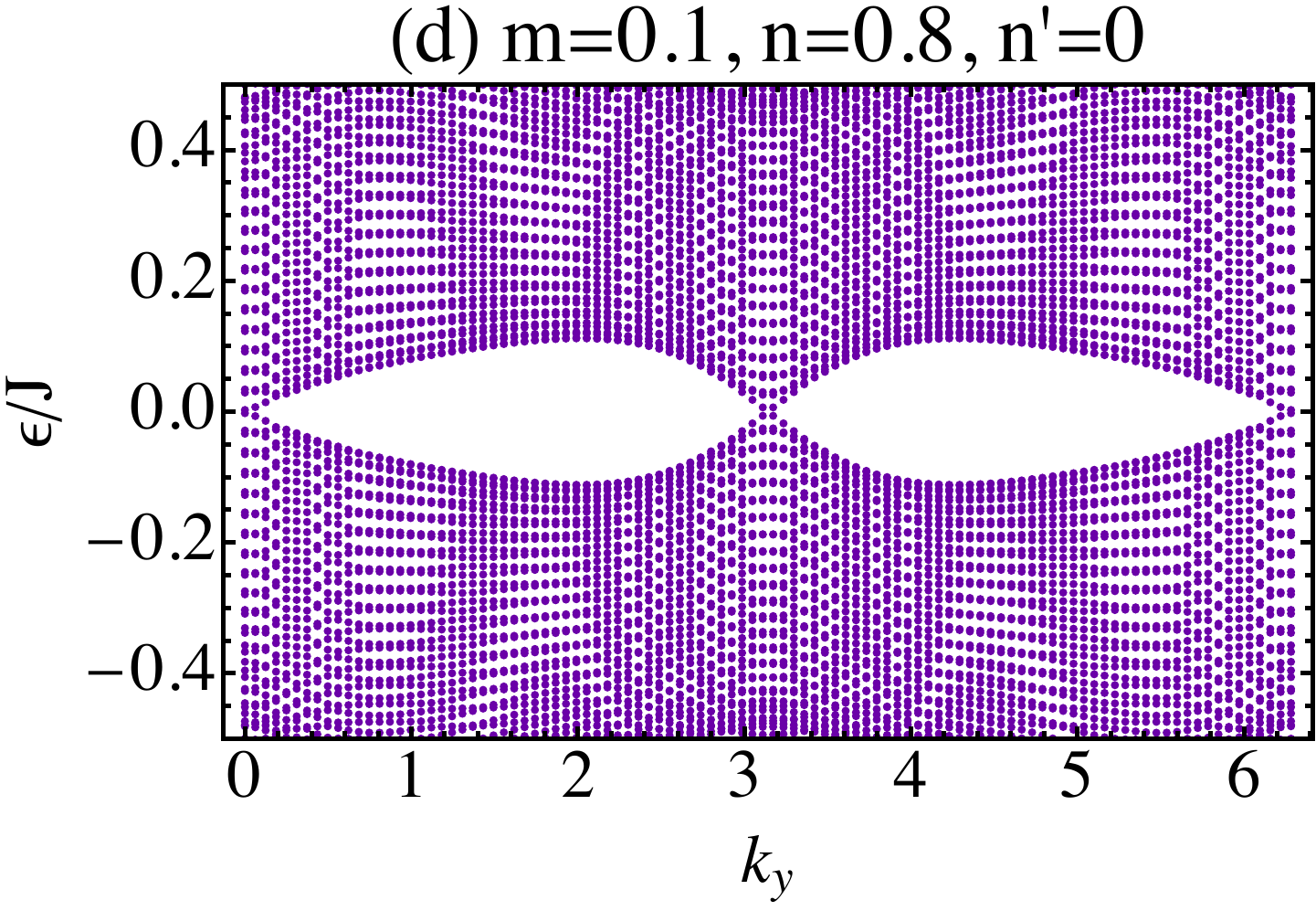}
\caption{Bandstructures for two points in the phase diagrams along $x$ and $y$ edges. The top panels (a,b) are for the extrinsic HOTI in the flat band phase, see Fig.\ \ref{fig:phasee}. The bottom panels (c,d) are for the intrinsic HOTI in the flat band phase, see Fig.\ \ref{fig:phasei}. (a,b) have fully bulk gap protected one dimensional flat bands along both edges (c,d) have one dimensional flat bands only along $k_x$, which appear between gap closings at specific momenta.} 
\label{fig:bs}
\end{figure}

The flat band phase which can be seen in Figs.\ \ref{fig:phasei} and \ref{fig:phasee} is different for the extrinsic and intrinsic HOTIs. In the extrinsic case $n'=n$ when there are no crystalline symmetries there is a robust flat band of one dimensional edge states with a bulk gap, see Fig.~\ref{fig:bs}(a,b). In the intrinsic case $n'=0$ when there {\it are} crystalline symmetries there are flat bands of one dimensional edge states with a bulk gap between nodal point at which the gap closes for particular momenta, see Fig.~\ref{fig:bs}(c,d). However these edge modes exist only along the $k_x$ direction suggesting a form of weak topology.

\section{Supplemental results on DQPTs}\label{app:dqpt}

\begin{figure}
\includegraphics[width=0.49\columnwidth]{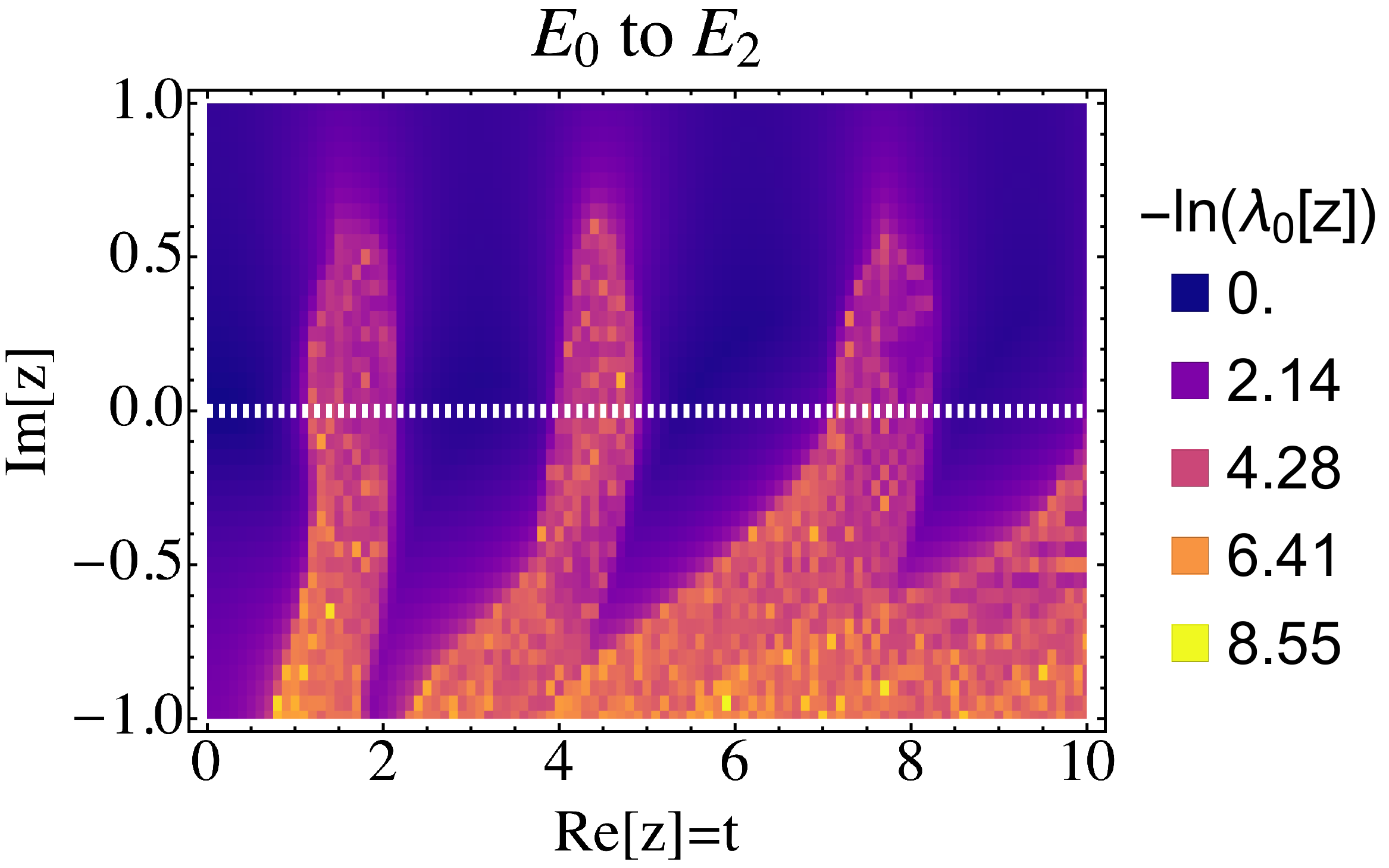}
\includegraphics[width=0.49\columnwidth]{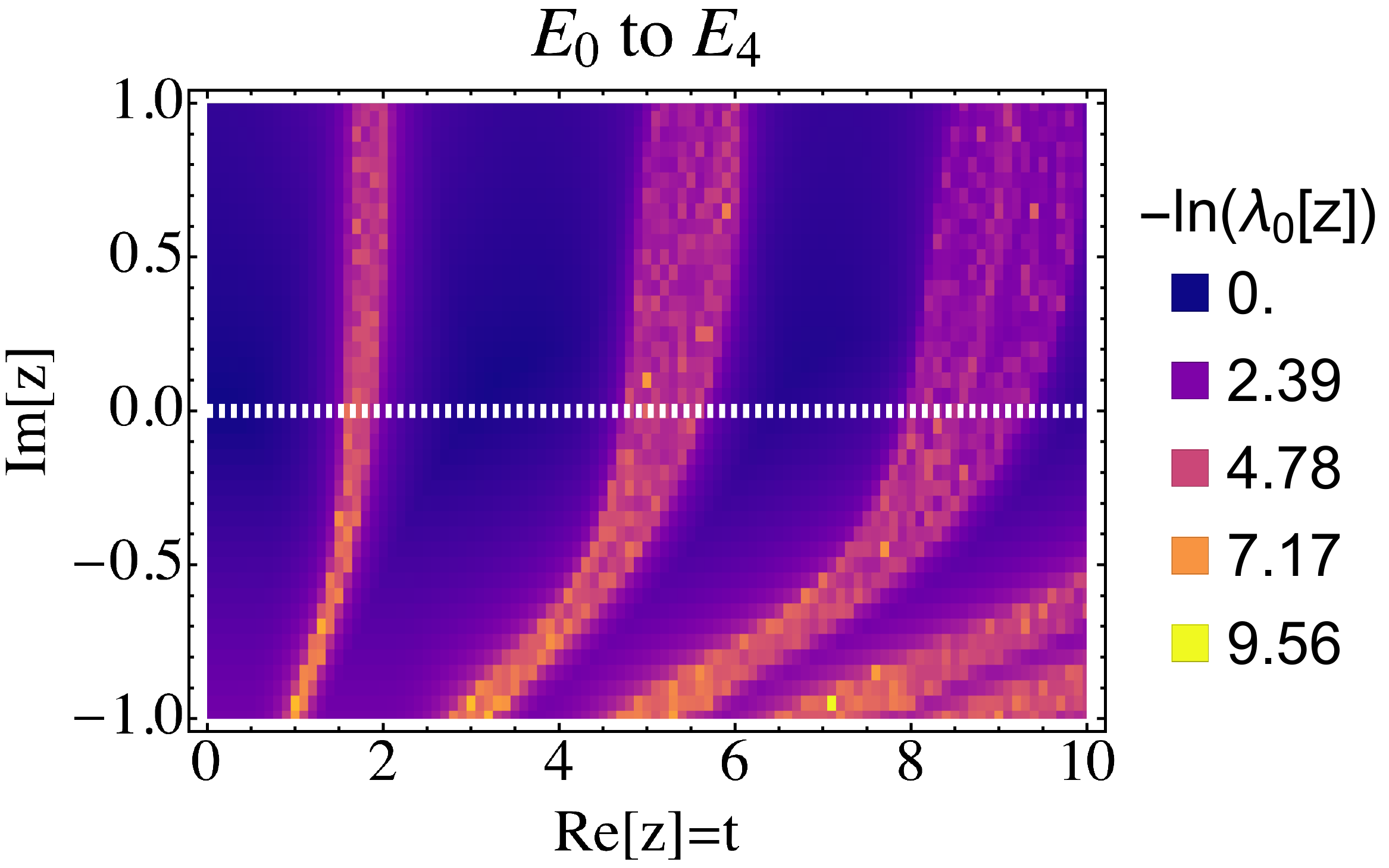}
\caption{Here we plot the proxy for the Fisher zeroes for two quenches, as labelled on the panels. Calculated for a system of size $N=202\times202$. In the thermodynamic limit Fisher zeroes would correspond to $-\ln[\lambda_0(z)]\to\infty$. Here we show results for a wider range of $z$, to make clear the aperiodicity (for the quench $E_0\to E_2$) and periodicity (for the quench $E_0\to E_4$) of the DQPTs along the real axis, i.e.~time.}
\label{fig:fishlong}
\end{figure}

In Fig.\ \ref{fig:fishlong} the Fisher zeroes are shown for longer ranges of time, equivalently the real part of $z$. The aperiodicity (for the quench $E_0\to E_2$) and periodicity (for the quench $E_0\to E_4$) of the DQPTs are clearly visible.

\begin{figure*}
\includegraphics[width=0.95\columnwidth]{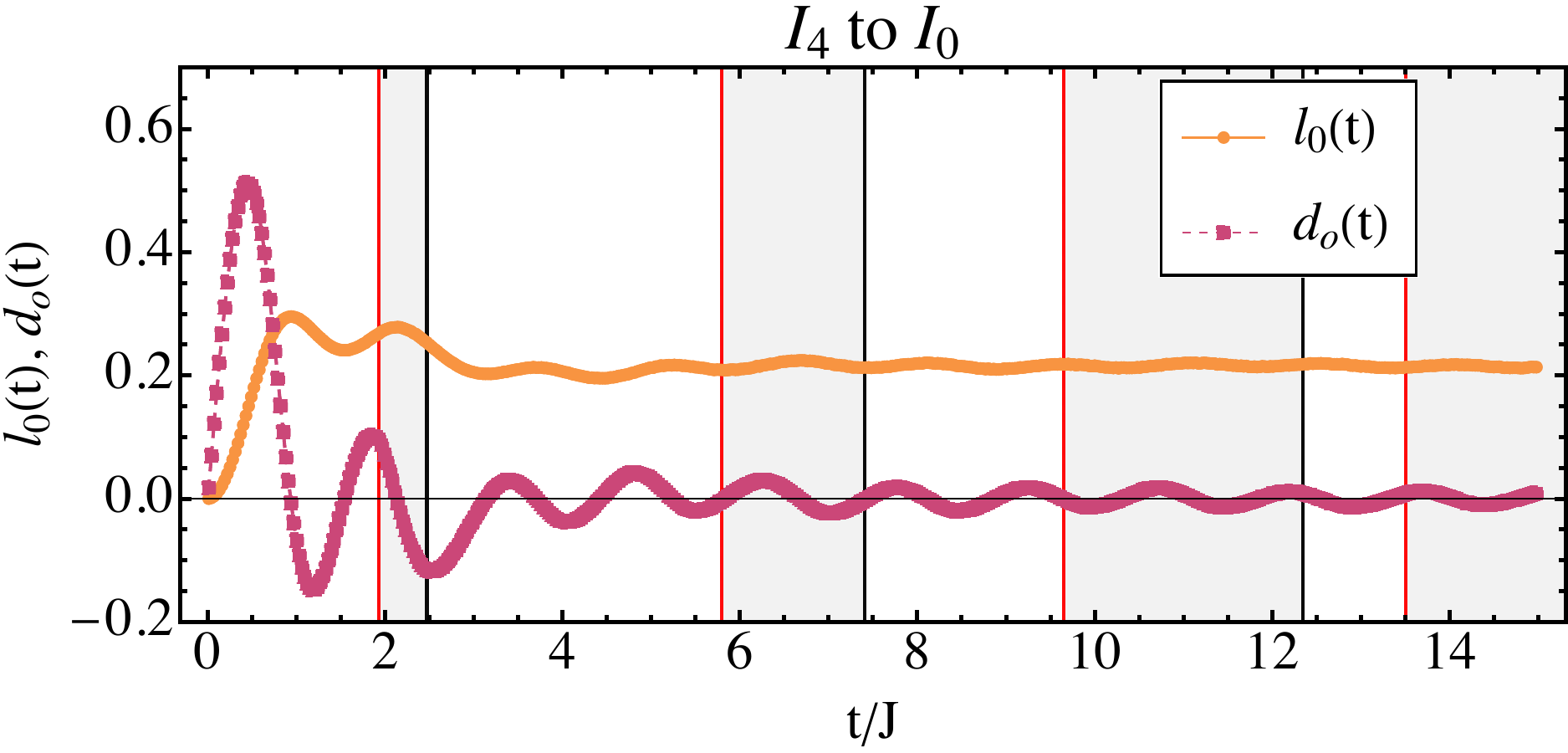}
\includegraphics[width=0.95\columnwidth]{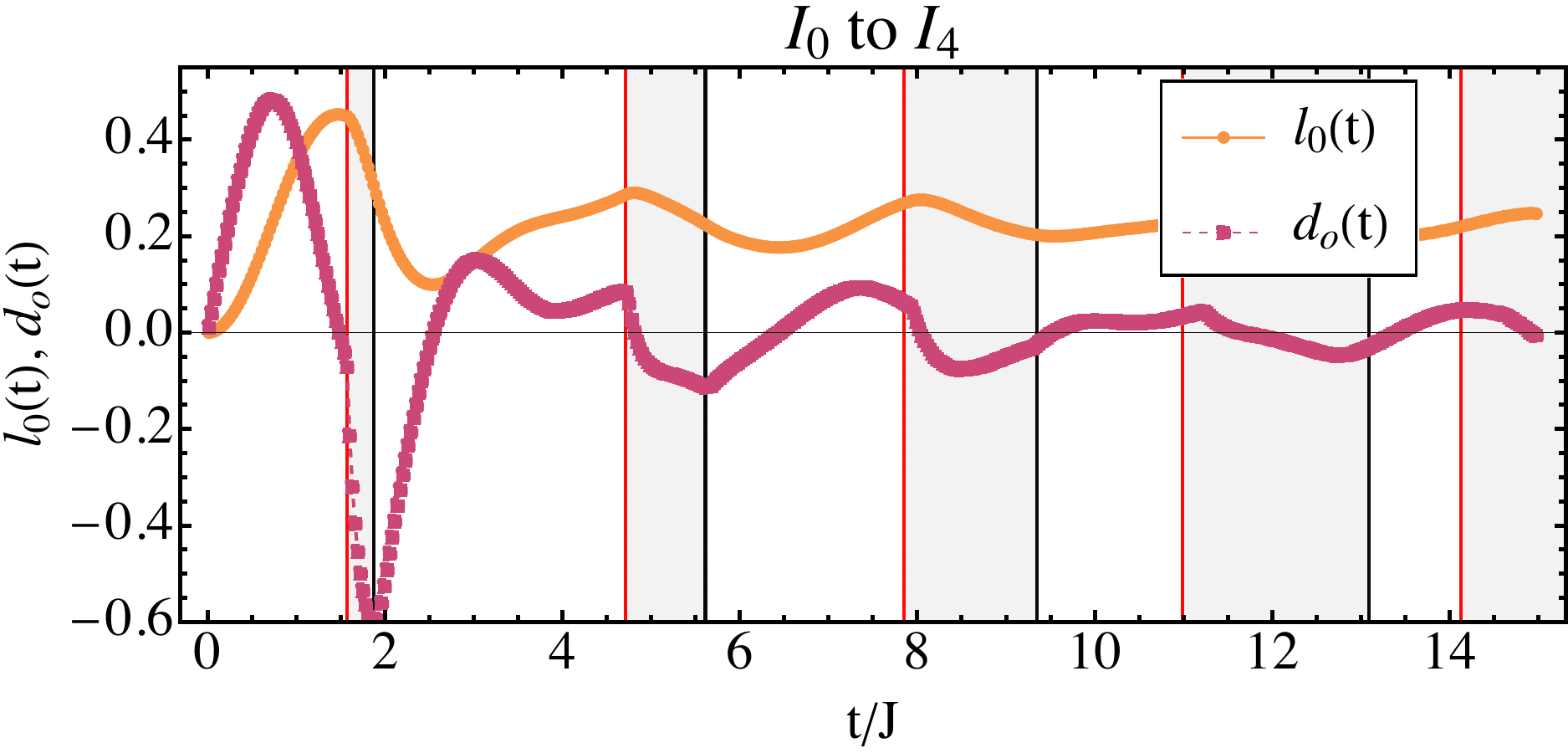}\\
\includegraphics[width=0.95\columnwidth]{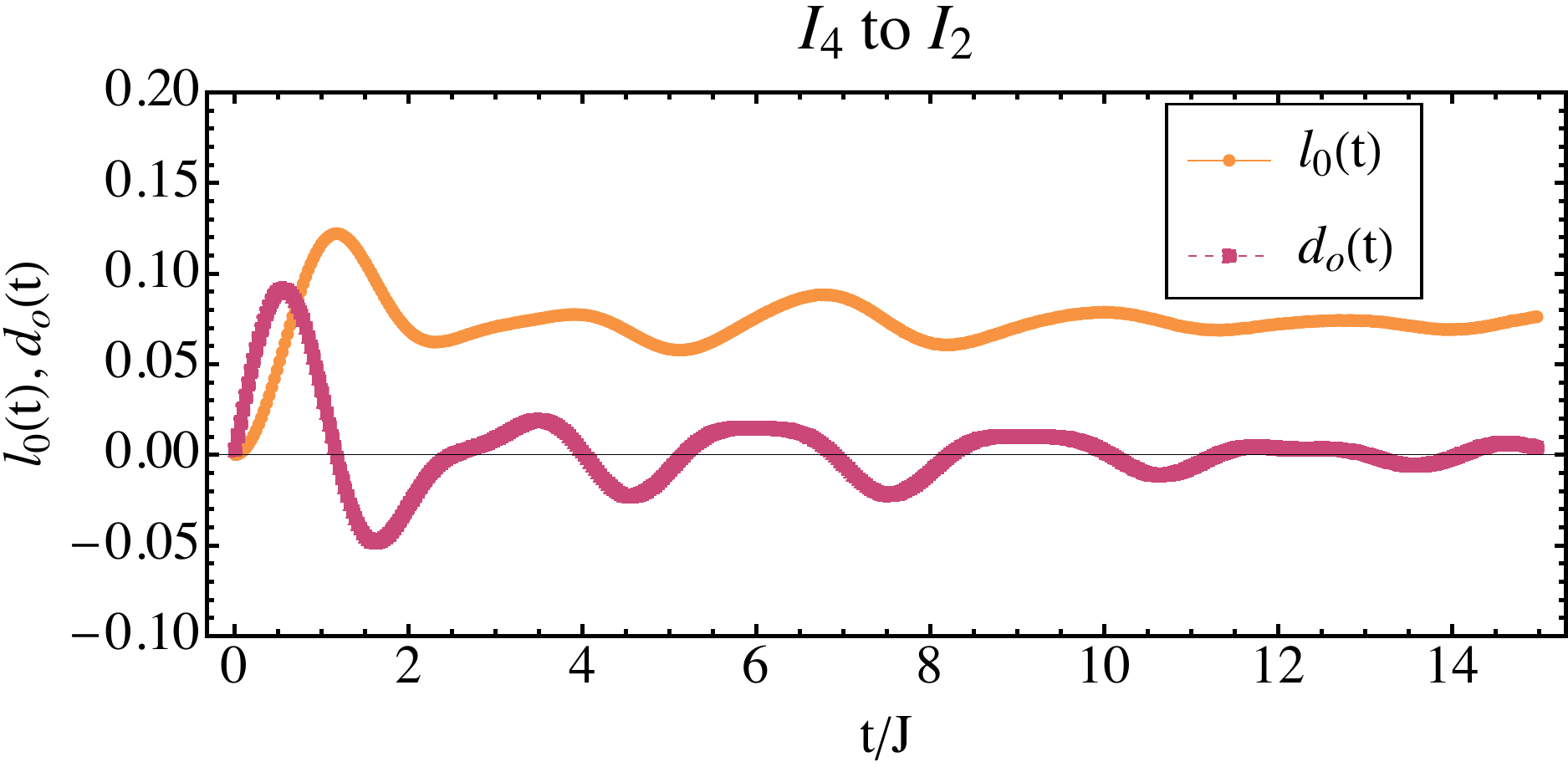}
\includegraphics[width=0.95\columnwidth]{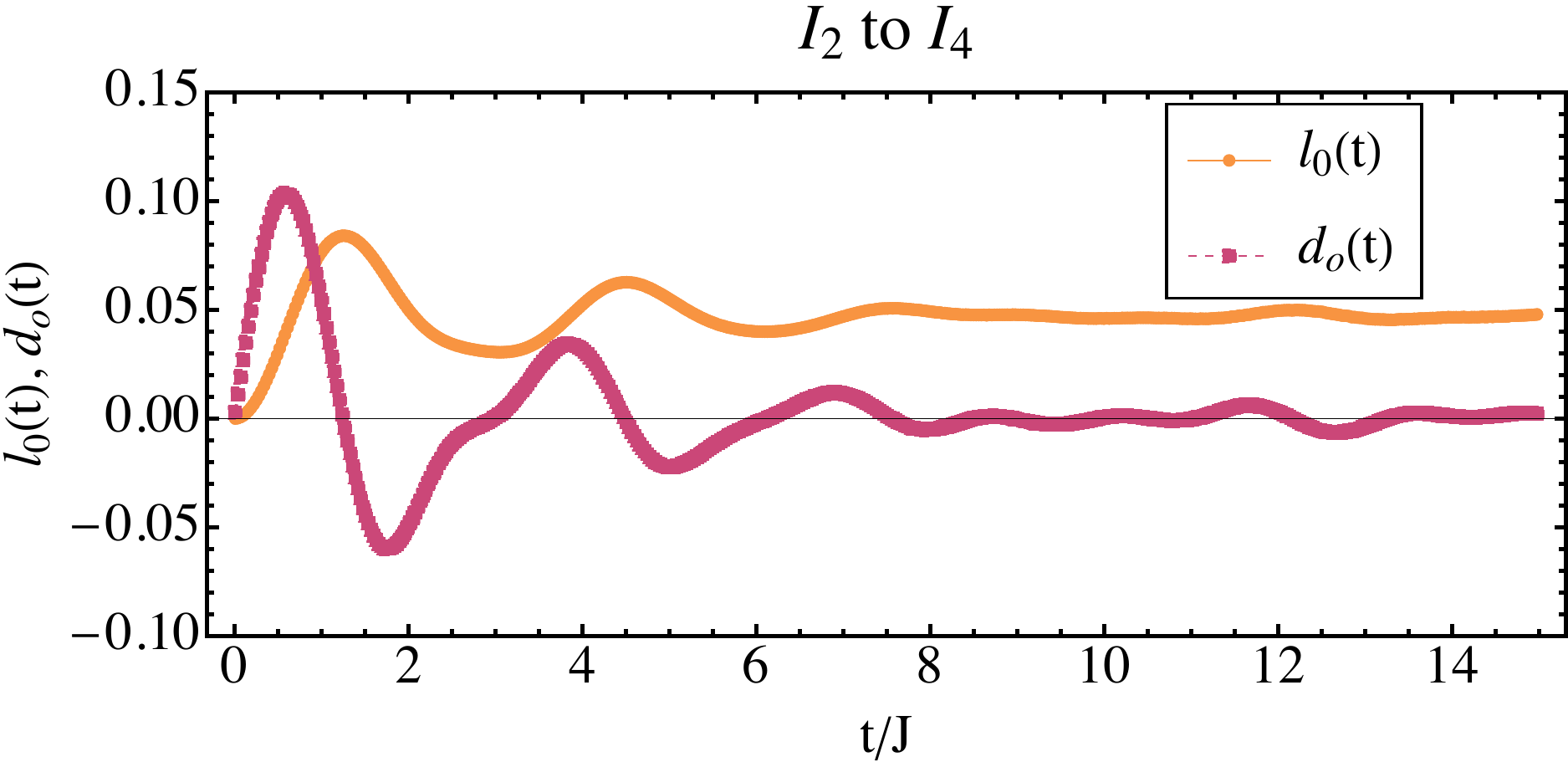}\\
\includegraphics[width=0.95\columnwidth]{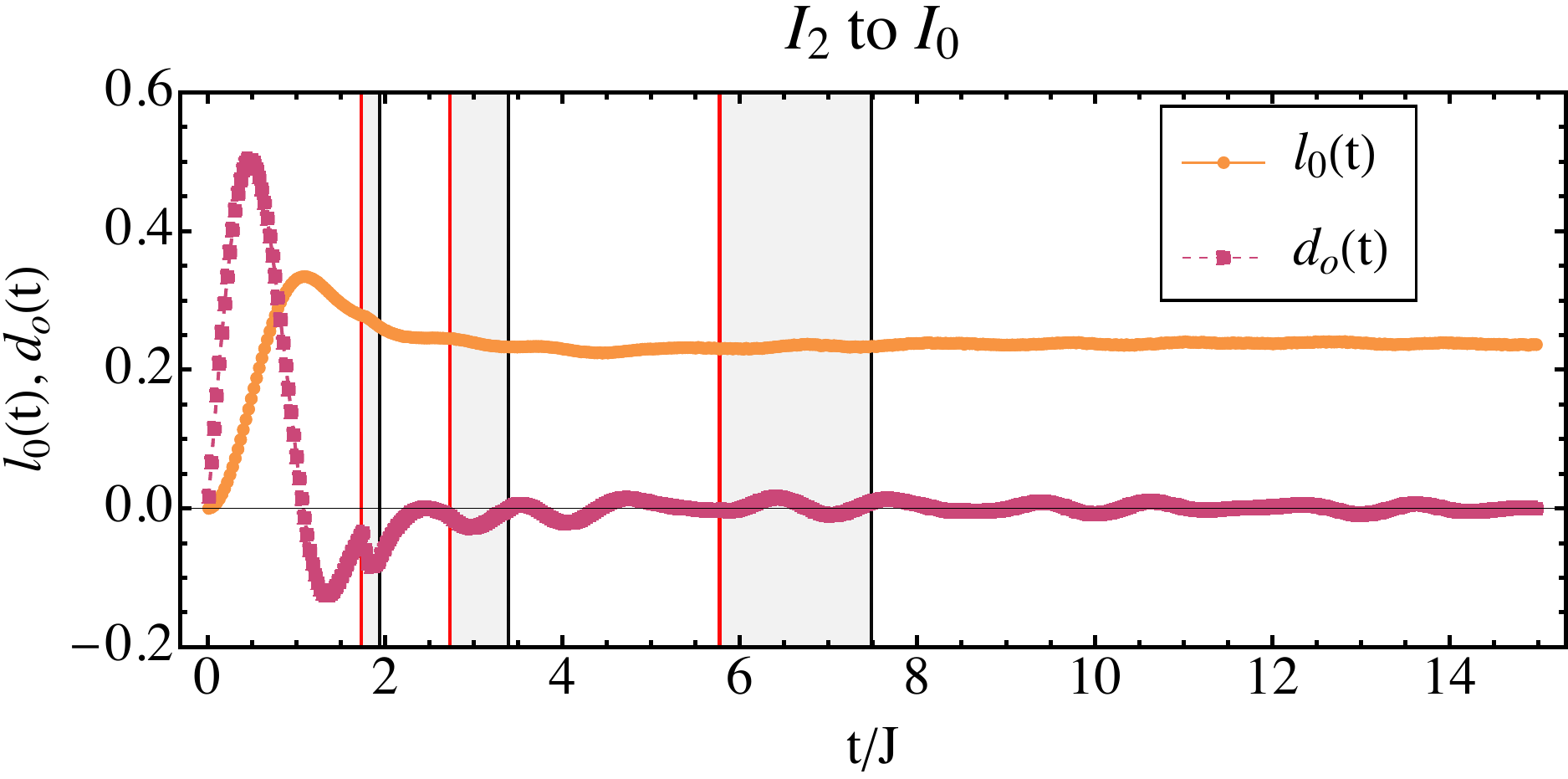}
\includegraphics[width=0.95\columnwidth]{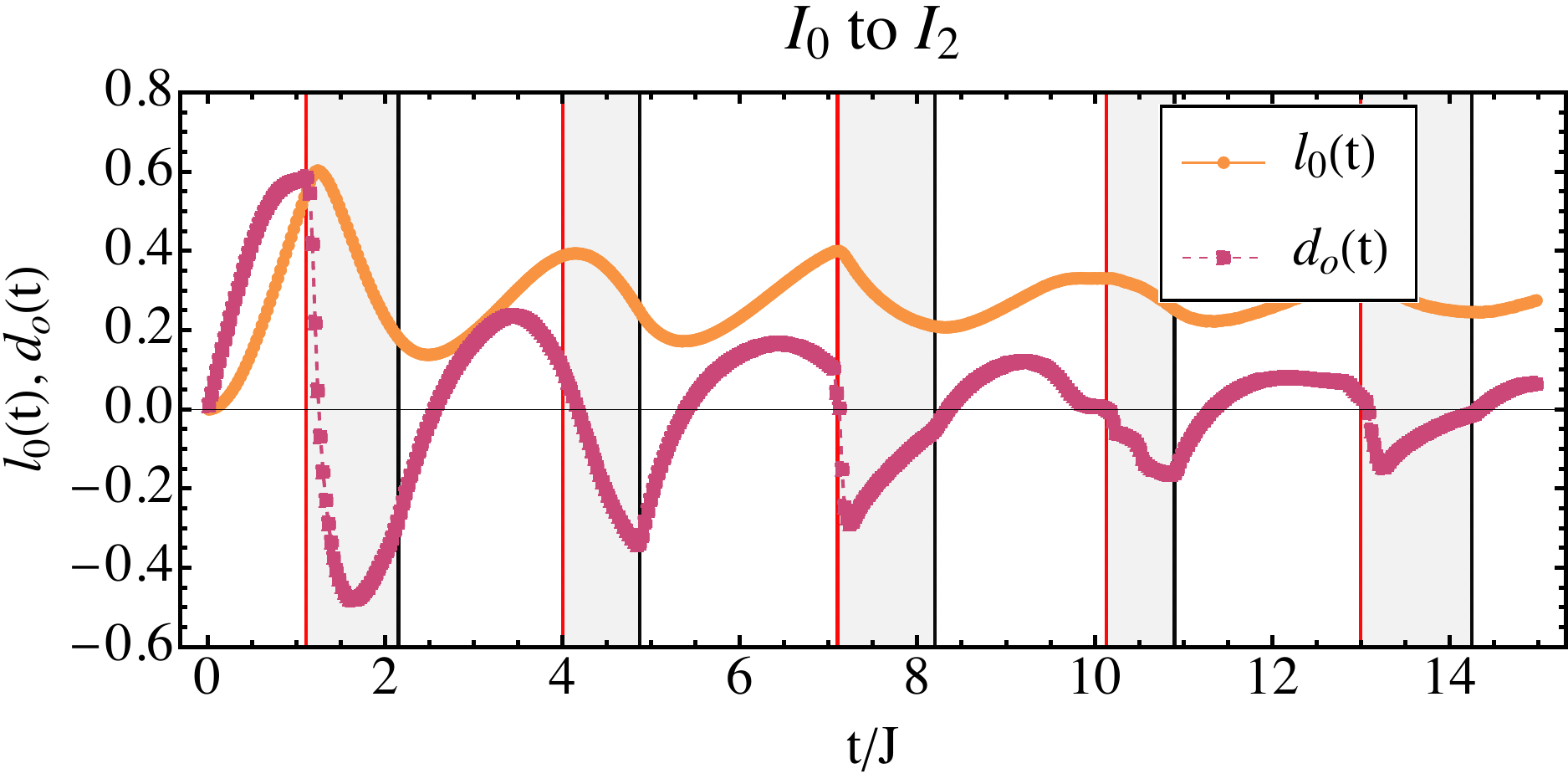}
\caption{The return rate $l_0(t)$ and its derivative $d_0(t)$ for several quenches of the intrinsic HOTI. The critical regions are shown in gray with $t_{c1}$ a red line and $t_{c1}$ a black line, where discontinuities in $d_0(t)$ are visible. These critical times are calculated from the eigenvalues of $M(t)$ at a system size of $N=202\times202$. For the quench $I_2\to I_0$ the critical regions are aperiodic and further regions could not be satisfactorily identified. For this model only quenches between the different topologically non-trivial phases did not result in DQPTs.}
\label{fig:brri}
\end{figure*}

Fig.\ \ref{fig:brri} gives a more comprehensive set of results for the return rate and its derivative for quenches in the intrinsic HOTI. All quenches except those between the different topologically non-trivial phases result in DQPTs. We did not find any DQPTs for quenches within any phase, though this can not be ruled out.

\begin{figure}
\includegraphics[width=0.95\columnwidth]{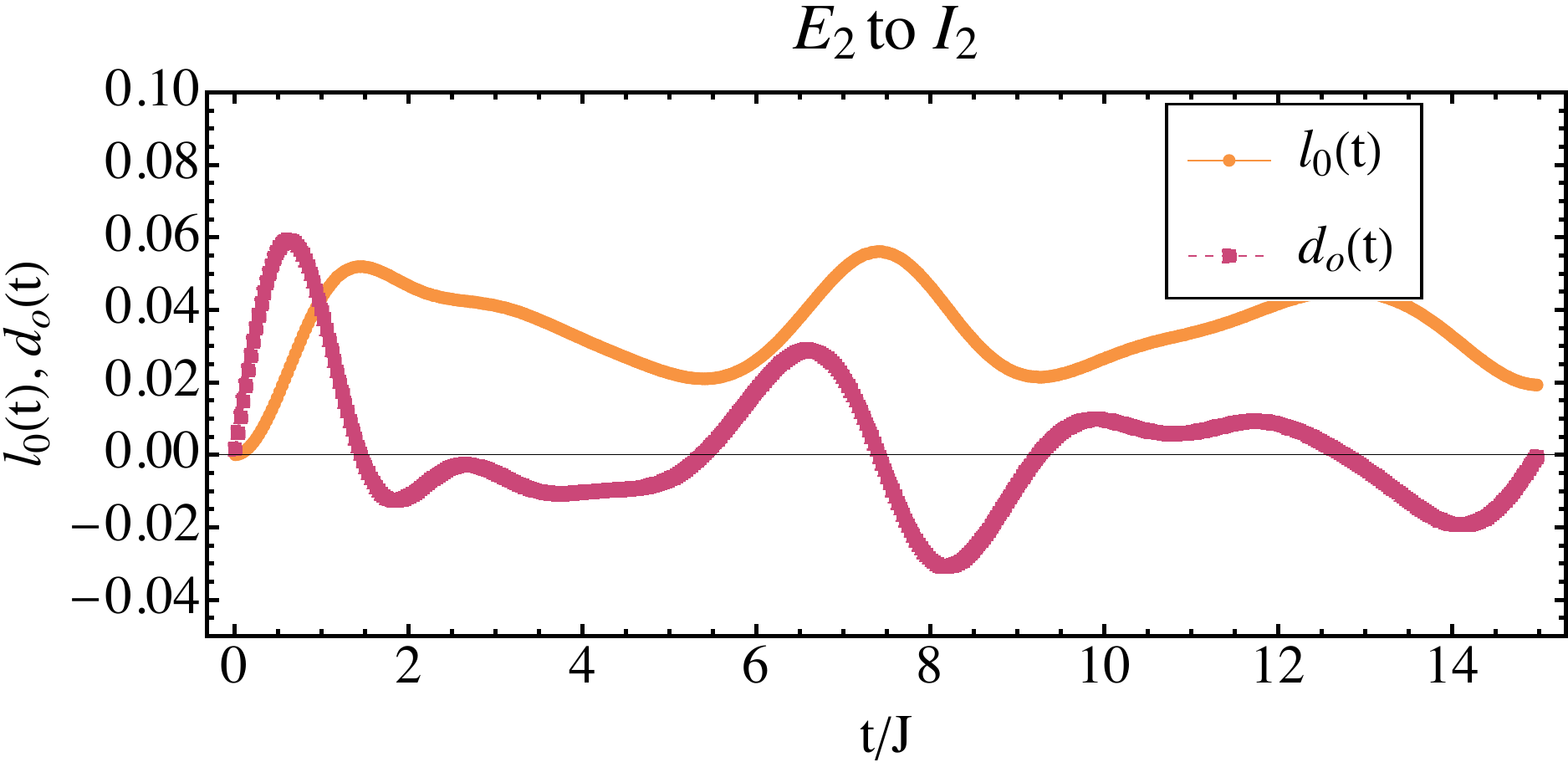}\\
\includegraphics[width=0.95\columnwidth]{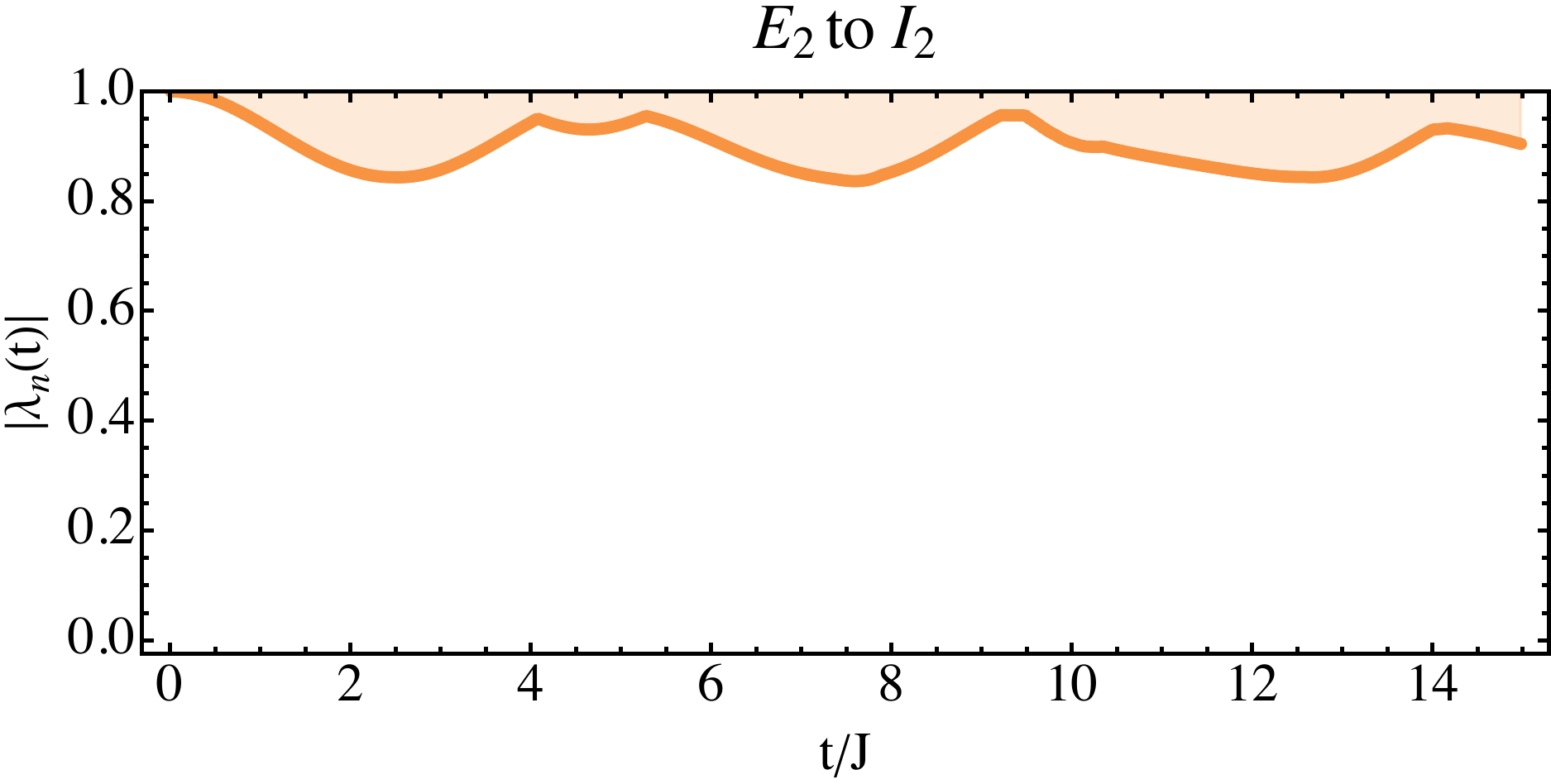}
\caption{Top panel: The return rate $l_0(t)$ and its derivative $d_0(t)$. Bottom panel: A plot of the absolute values of the smallest eigenvalues of $M(t)$, $|\lambda_n(t)|$. Shown is the lowest eigenvalue for a periodic system of size $N=202\times 202$, orange dots. The shaded region is the region where eigenvalues of $M(t)$ exist for the bulk system. No DQPTs can be sen for this quench, $E_2\to I_2$.}
\label{fig:brrei}
\end{figure}

Finally for completeness Fig.\ \ref{fig:brrei} gives one example of a quench between the extrinsic and intrinsic HOTI. This does not cross a critical line, but does restore a symmetry. No DQPTs can be seen. Similar results were found for all quenches $I_z\leftrightarrow E_z$ with $z=0,2,4$.

%

\end{document}